\documentclass[fleqn,10pt]{wlscirep}
\usepackage[utf8]{inputenc}
\usepackage[T1]{fontenc}

\usepackage{amsmath,amsthm,amssymb,amsfonts,stmaryrd}
\usepackage{graphicx}
\usepackage{color}
\usepackage{arydshln}	% Dotted lines in arrays
\usepackage{adjustbox}
\usepackage{hyperref}
\AtBeginDocument{}
\usepackage{chemformula}
\usepackage{multirow}
\usepackage{booktabs}
\usepackage{siunitx}
\usepackage{subcaption}
\usepackage{afterpage}
\usepackage{placeins}

\hypersetup{
  colorlinks   = true, %Colours links instead of ugly boxes
  urlcolor     = blue, %Colour for external hyperlinks
  linkcolor    = blue, %Colour of internal links
  citecolor   = red %Colour of citations
}
\usepackage{cleveref}
\usepackage{multirow}
\usepackage{comment}

\DeclareSIPrefix\micro{\text{\textmu}}{-3}

\usepackage{amsmath}
\usepackage{amsfonts}
\usepackage{amssymb}
\usepackage{mathtools}
\usepackage{bm}

\newcommand{\p}{\mathbf{p}}

\newcommand{\bmu}{\boldsymbol{\mu}}
\newcommand{\bsigma}{\boldsymbol{\sigma}}
\newcommand{\bzero}{\boldsymbol{0}}
\newcommand{\bI}{\boldsymbol{I}}

\DeclarePairedDelimiterX{\infdivx}[2]{(}{)}{%
  #1\;\delimsize\|\;#2%
}


\title{Learning robust parameter inference and density reconstruction in flyer plate impact experiments}

\author[1]{Evan Bell}
\author[1,*]{Daniel A. Serino}
\author[1]{Ben S. Southworth}
\author[2]{Trevor Wilcox}
\author[1]{Marc L. Klasky}

\affil[1]{Theoretical Division, Los Alamos National Laboratory, P.O. Box 1663, Los Alamos, NM 87545 U.S.}

\affil[2]{Theoretical Design Division, Los Alamos National Laboratory, P.O. Box 1663, Los Alamos, NM 87545 U.S.}

\affil[*]{dserino@lanl.gov}

\begin{abstract}
    {Estimating physical parameters or material properties from experimental observations is a common objective in many areas of physics and material science. In many experiments, especially in shock physics, radiography is the primary means of observing the system of interest. However, radiography does not provide direct access to key state variables, such as density, which prevents the application of traditional parameter estimation approaches. Here we focus on flyer plate impact experiments on porous materials, and resolving the underlying parameterized equation of state (EoS) and crush porosity model parameters given radiographic observation(s). We use machine learning as a tool to demonstrate with high confidence that using only high impact velocity data does not provide sufficient information to accurately infer both EoS and crush model parameters, even with fully resolved density fields or a dynamic sequence of images. We thus propose an observable data set consisting of low \emph{and} high impact velocity experiments/simulations that capture different regimes of compaction and shock propagation, and proceed to introduce a generative machine learning approach which produces a posterior distribution of physical parameters directly from radiographs. We demonstrate the effectiveness of the approach in estimating parameters from simulated flyer plate impact experiments, and show that the obtained estimates of EoS and crush model parameters can then be used in hydrodynamic simulations to obtain accurate and physically admissible density reconstructions. Finally, we examine the robustness of the approach to model mismatches, and find that the learned approach can provide useful parameter estimates in the presence of out-of-distribution radiographic noise and previously unseen physics, thereby promoting a potential breakthrough in estimating material properties from experimental radiographic images.}
\end{abstract}

\begin{document}
\maketitle
% \tableofcontents

\allowdisplaybreaks

\section{Introduction}
\label{sec:intro}

\subsection{Background and motivation}
The dynamic compaction of granular porous materials is of fundamental interest within a variety of scientific disciplines, including geo- and astro-physics, shock physics, energetic material dynamics, and high energy density physics \cite{jorgensen1989using,collins2019planetary,sharma2023effect,li2020hotspot,tikhonchuk2024physics}. For example, rapid astrophysical compaction processes, such as impact cratering, are significantly affected by the dynamic characteristics of shock compaction in the constituent materials\cite{housen2003impact,nakamura2017impact}. In investigations of  material properties, predictive modeling is complicated by uncertainty in parameters that characterize various aspects of the system, such as the equation of state (EoS) and accompanying crush model, as well as the initial conditions. Consequently, dynamic experimentation plays a crucial role in calibrating models to improve simulations of hydrodynamic behavior and facilitate the discovery of material properties.  

Historically, material properties including EoS and constitutive relationships (such as material strength in shock physics and material science) have been investigated via an impulse-response approach, where the velocity trace response of a material specimen to an impulse is measured using velocity interferometry\cite{barker1972laser,malone2006overview,mccoy2017lagrangian,reinhart2001equation}. Indeed, the development of laser interferometry has enabled the time-resolved measurement of the velocity of a reflecting surface\cite{barker1972laser}, which has allowed for the measurement of the free surface and window interface velocities in dynamic compression experiments\cite{mccoy2017lagrangian}. These measurements have yielded valuable data on compressive behavior and strength of materials during both shock compression and release in dynamic experiments\cite{asay1978self,lipkin1977reshock,brown2013extracting,mcmcflyer}. However, in extreme environments, this technique may not be feasible\cite{brown2013extracting,barnes1974taylor,colvin2003model,barton2011multiscale}. In this case, radiography provides the main experimental diagnostic used to probe the material properties. For example, radiography has been used in gas gun-driven flyer plate impact experiments to make experimental Hugoniot measurements\cite{hugoniotxray}. % and to validate theoretical models of compaction for granular rock samples\cite{shockxray}.  
However, calibrating physical models using experimental radiographic data from dynamic imaging experiments presents unique challenges\cite{serino2024learning}.
In particular, radiography does not give a direct measurement of key state variables, such as density, temperature, pressure, or even an interface velocity. Consequently, this problem largely precludes the direct application of many common parameter estimation techniques that use computational models of the underlying physics, such as Bayesian optimization, Markov chain Monte Carlo (MCMC) sampling, or PDE-constrained optimization. Furthermore, accurately extracting the primary state variable, density, from noisy radiographic measurements continues to be a challenge, even using modern image reconstruction methods, such as model-based image reconstruction (MBIR) or statistical image reconstruction (SIR) due to the presence of scattered radiation, energy beam spectrum uncertainty, beam spectral effects, and unknown system noise\cite{ravishankar2019image,fessler2010model,elbakri2002statistical,espy2021spectral,espy2016wide,sauer2002local,mccann2021local}. 

Recently, machine learning (ML) approaches have been applied to radiographic reconstruction to address or circumvent some of these challenges\cite{xu2024swap,huang2022physics,serino2024reconstructing,hossain2022high,lahiri2023sparse}, with many of these ML architectures outperforming MBIR methods by a large margin at a specified degradation level\cite{huang2022physics,ravishankar2019image}. Yet, it still remains unclear whether the density reconstructions produced by these approaches are accurate enough to enable parameter estimation by traditional methods\cite{shockdensityrecon}. Moreover, even if the obtained reconstructions are sufficiently accurate, applying traditional parameter estimation methods may still prove challenging. For example, MCMC methods generally scale poorly to problems where the parameter space has a moderate or high dimension\cite{mcmcreview}, and PDE-constrained optimization typically requires differentiable simulators which are often not available. Consequently, we develop an ML-based inverse approach for extracting parameters that govern the behavior of porous materials. In particular, our approach enables simultaneously estimating both crush model and Mie-Grüneisen EoS parameters under dynamic loading conditions using radiographic projections. In general, the inverse approach has several advantages relative to the more experimentally intensive direct approach\cite{pierron2021towards,tariq2025inverse}. In this context, the direct method assumes an EoS and involves determining the crush model parameters from repeated shock physics experiments. Indeed, there is a notable surge in research aimed at reducing the reliance on traditional mechanical tests by utilizing a combination of inverse methodologies, heterogeneous tests, and full-field measurements, often referred to as Material Testing 2.0\cite{pierron2021towards}. The inverse
method specifically focuses on identifying parameters of simulations codes that are consistent with a set of experimental observations. The use of these methods in  computational mechanics has progressed rapidly in recent years\cite{tariq2025inverse,prates2016inverse,markiewicz2017review,pottier2013inelastic}. Furthermore, the direct approach ignores possible correlations between EoS and the crush-model parameters, and it has been found that the inverse method utilizing optimization algorithms can effectively estimate parameters that are challenging to measure directly\cite{bonfiglio2013inversion}. Following inference of EoS and crush parameters from radiographic image(s), one can immediately plug these parameters into a high-fidelty forward model to yield a physically admissible density reconstruction at any time.

\begin{figure}[!t]
    \centering
    \includegraphics[width=0.8\linewidth]{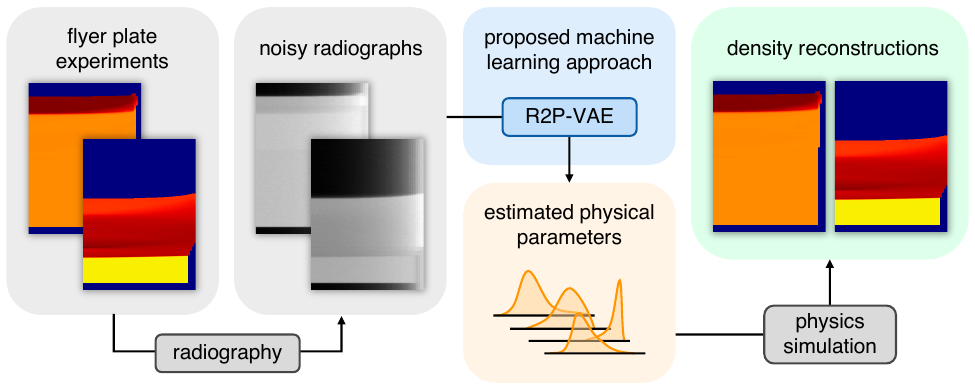}
    \caption{Overview of the proposed approach for parameter estimation and density reconstruction from radiographs acquired in flyer plate impact experiments. The radiographs-to-parameters variational autoencoder (R2P-VAE) produces distributions of physical parameters, which can be used in a hydrodynamic solver to obtain density fields and other state variables.}
    \label{fig:flowchart}
\end{figure}

\begin{comment}
Finally, we remark that ML has previously been successfully utilized to investigate the properties of porous materials. For example, convolutional neural networks have been utilized to predict petrophysical properties of rocks, such as specific surface area and average pore size, using 2D X-ray images\cite{alqahtani2020machine}, demonstrating the potential of machine learning in characterizing porous media. Deep learning models have also been used to predict the permeability of porous media from either images or point clouds, offering a faster and more computationally efficient alternatives to traditional simulation methods\cite{takbiri2022deep, kashefi2021point}. Machine learning has also been used to predict material properties in other contexts, such as predicting the mechanical properties of metal-organic frameworks\cite{moosavi2020role}, or predicting the parameters of the Johnson-Champoux-Allard model of fibrous materials given geometric parameters of the material\cite{yi2024machine}.
\end{comment}

In the remainder of this work, we present the details of a generative ML approach to simultaneously infer EoS and $P-\alpha$ crush parameters from radiographic images, {using simulated flyer plate impact experiments as a test problem.} We summarize the proposed approach graphically in \autoref{fig:flowchart}. We also seek to determine sufficient conditions with respect to the flyer plate velocities and number of discrete velocities sampled for accurate parameter inference across EoS and crush models. In \autoref{sec:problem}, we introduce the EoS and crush models for which will seek to determine parameters from radiographic images, describe the flyer plate test problem and details of the data generation, and present our proposed ML framework for parameter estimation and density reconstruction.  {In \autoref{sec:parameter_estimation}, we present our results on parameter estimation. We begin by considering the hypothetical case where the underlying density fields are accessible for parameter estimation. We use ML as a tool to demonstrate that even in this ideal scenario, observations from multiple flyer plate experiments with impact velocities in distinctly different regimes are required for accurate inference of all of relevant physical parameters in EoS and crush models. These results inform the subsequent design of the central experiments, in which we perform parameter estimation from radiographic measurements which are corrupted with noise and scatter. Then, in \autoref{sec:density_recon}, these parameter estimates are used to perform density reconstruction by passing them through a hydrodynamic solver. In this section, we also demonstrate the efficacy of our approach under two different model mismatch scenarios. 
First, we examine a mismatch in the model of the radiographic system by applying our trained parameter estimation network to radiographs corrupted with out-of-distribution noise. Second, we perform parameter estimation and density reconstruction for a test case where the EoS is parameterized using a fundamentally different model than the one used for training. Finally, in \autoref{sec:discussion} we discuss the simulation results and conclusions drawn from the numerical experiments.}

\section{Problem and methods}
\label{sec:problem}

\subsection{Porous materials and flyer plate impact experiments}
\label{sec:modeling_porus}

\begin{figure}[!tbh]
    \centering
    \includegraphics[width=0.8\textwidth]{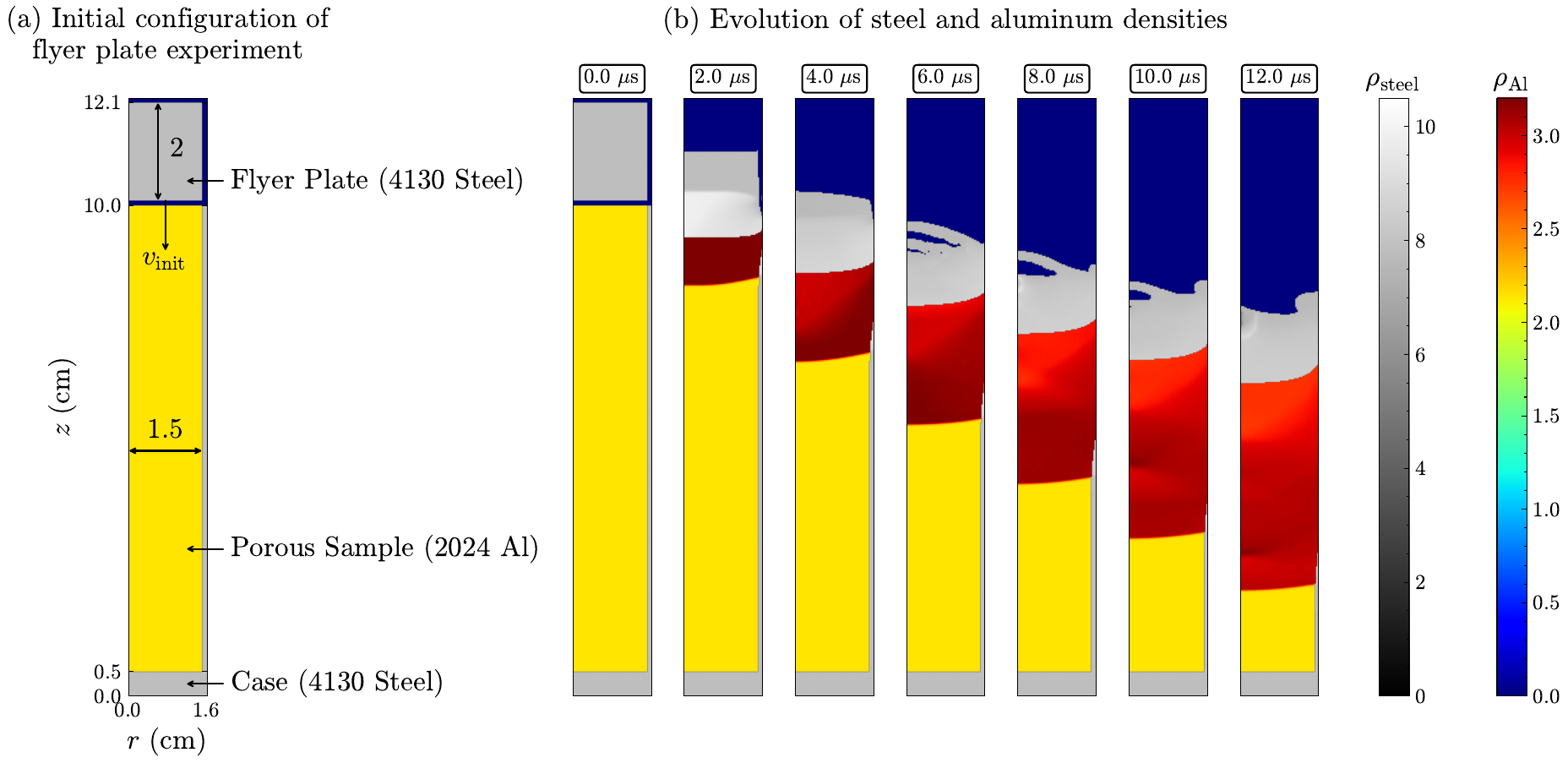}
    \caption{Flyer plate experiment set-up and example impact. (a) initial geometry of the flyer plate experiment, (b) evolution of densities of the steel and aluminum over 12 $\si{\us}$. Densities are in units of g/cm$^3$. The impact shown in (b) is for the high impact velocity experiment, with $v_{\rm init}=5\cdot10^5~\si{cm/s}$.}
    \label{fig:initial}
\end{figure}

As a model problem of flyer plate impact experiments,
we consider a flyer plate impacting a porous material sample
based on a 1974 experiment at 
Sandia National Laboratory~\cite{butcher74,drumheller78} designed to facilitate calibration of parameters for a porosity model of 2024 aluminum. More specifically, we consider the impact of a 4130 steel flyer plate on a porous 2024 aluminum sample enclosed in a 4130 steel case. We assume azimuthal symmetry in the 3D cylindrical domain, and the resulting 2D computational domain can be described in $(r,z)$ coordinates via
$(r, z) \in \Omega = [0, 1.6{\rm cm}] \times [0, 12.2 {\rm cm}]$.
The initial geometry is detailed in ~\autoref{tab:initgeom} 
and illustrated in ~\autoref{fig:initial}(a).
The flyer plate is given an initial downward velocity, $v_{\rm init}$.
The steel flyer plate then hits the aluminum and compresses it, crushing the pores in the material and generating heat. A strong front shock wave is generated, which propagates through the 
material. A secondary shock also emerges behind the primary shock. See ~\autoref{fig:initial}(b) for an illustration of the evolution of the densities of the steel and aluminum during the impact.
The dynamic evolution, compaction, followed by shock compression and propagation are defined by the multimaterial compressible Euler (also referred to herein as hydrodynamics) equations subject to EoS, material strength and porosity models. Here, the steel is modeled using a Mie-Gr\"uneisen model for the EoS and an elastic perfect plasticity based on Von Mises yield surface (EPPVM)~\cite{taylor2006introduction} model for the material strength.
The aluminum sample is modeled using a Mie-Gr\"uneisen model for the EoS
coupled with a $P-\alpha$ porosity model. 

\begin{table}[htb]
\centering
\begin{tabular}{cccccc}
\toprule
component & material & $r_{\min}$ (cm) & $r_{\max}$ (cm) & $z_{\min}$ (cm) & $z_{\max}$ (cm) \\
\midrule
flyer plate & 4130 steel & 0 & 1.5 & 10.1 & 12.1 \\
case wall & 4130 steel & 1.5 & 1.6 & 0 & 10 \\
case bottom & 4130 steel & 0 & 1.6 & 0 & 0.5 \\
porous material & 2024 aluminum & 0 & 1.5 & 0.5 & 10 \\
\bottomrule
\end{tabular}
\caption{Initial geometry the of flyer plate experiment.}
\label{tab:initgeom}
\end{table}

Let $\rho, T, P, E$ be the density, temperature, pressure, and internal
energy of a material. These four variables define a thermodynamic phase space. 
The Mie-Gr\"uneisen equation of state (EoS) model relates $(P, E)$ as a function  of $(\rho, T)$ using\cite{cth92,rice58}
\begin{align}
    P(\rho, T) = P_R(\rho) + \Gamma_0 \rho_0 (E(\rho, T) - E_R(\rho)),
    \qquad
    E(\rho, T) = E_R(\rho) + C_V (T - T_R(\rho)),
    \label{eq:mgrun}
\end{align}
where 
$\Gamma_0$ is the Gr\"uneisen parameter,
$C_V$ is the specific heat, and
$P_R, E_R, T_R$ are reference 
curves on an isentrope that intercepts the
Hugoniot through the initial state
$(\rho_0, T_0, P_0, E_0).$
The reference curves are defined to be
$T_R(\rho) = T_0 \exp(\Gamma_0 \mu),$
$E_R(\rho) = E_0 + \frac{\mu}{\rho_0} P_0 + \frac{c_s^2 \mu^2 Y(\mu)}{2(1 - s \mu)},$
and $P_R(\rho) = \rho_0 \frac{dE_R}{d\mu}$,
where $\mu = 1 - \frac{\rho_0}{\rho}$,
$c_s$ is the sound speed of a Hugoniot fit,
$s$ is a constant, and  $Y(\mu) = \sum_{k=0}^{\infty} a_k \mu^{k}$,
where the coefficients are defined by the recursion
relation
$a_0 = 1$, 
$a_1 = \frac{s}{3}$, and
$a_k = \frac{1}{k+2} \left((\Gamma_0 + ks) a_{k-1} - \Gamma_0 s a_{k-2} \right)$ for $k \ge 2.$
In this paper, we utilize the CTH simulation code~\cite{cth92}
which uses only the first five coefficients, $a_0, \dots a_4$
to compute $Y_\mu$.
In summary, the Mie-Gr\"uneisen EoS model
can be defined in terms of the parameters
$(\rho_0, T_0, \Gamma_0, C_V, c_s, s)$.

The $P{\rm -}\alpha$ porosity model~\cite{Herman69} is a modification of the EoS to account for voids in porous materials. The $P{\rm -}\alpha$ model is a phenomenological approach that separates the volumetric response of a porous material into two components: pore collapse (compaction) and matrix compression\cite{Herman69,carroll1972suggested,menikoff2000equation}.  It introduces a distention parameter, $\alpha$, defined as the ratio of the specific volume of the porous material to that of the fully compacted solid matrix. This model assumes that the specific internal energy of the porous material is equivalent to that of the solid material at the same pressure and temperature, allowing for a simplified yet effective representation of the compaction behavior. The $P{\rm -}\alpha$ model has been implemented in various computational tools and has undergone several modifications to enhance its applicability\cite{mahon2015compaction,jutzi2008numerical,de2018low,wang2023modified}.
The distention parameter $\alpha$ is defined as the ratio
$\alpha = \frac{\rho_0}{\rho},$
where $\rho$ is the porous material density
and $\rho_0$ is the reference density of the material
without voids.
The model relates the EoS of the porous material 
to the EoS of the void-free material using
% \begin{align}
    $P(\rho, T, \alpha)
    = \frac{1}{\alpha} \bar{P}(\alpha \rho, T), \qquad 
    E(\rho, T, \alpha)
    = \bar{E}(\alpha \rho, T)$,
% \end{align}
%
where $(\bar{P}, \bar{E})$ are EoS relations
for the void-free material and can be modeled
using~\eqref{eq:mgrun}.
The distention parameter is a time evolving parameter
that is modeled using the system
% \begin{align}
    $\frac{d\alpha}{dt} = 
    \frac{d\alpha}{dP} \frac{dP}{dt}
    = \alpha_P \frac{dP}{dt}$,
% \end{align}
%
where the initial value is $\alpha_0 = \frac{\rho_0}{\rho_{p}}$, where $\rho_{p}$ is the initial porous
density. 
The model divides the behavior into two regions, 
including an elastic region which is reversible
and a compaction region which is irreversible.
Let $P_e$ is the maximum elastic pressure,
we define the maximum distention as a function
of pressure to be
% \begin{align}
    $\alpha_{\rm max}(P) =
    1 + (\alpha_0 - 1)
   \left(\frac{P_s - P}{P_s - P_e}\right)^{n},$
% \end{align}
%
where $P_s$ is the pressure for complete compaction $\left(\alpha_{\rm max}(P_s) = 1\right)$
and $n$ is an parameter.
$\alpha_P$ is modeled as
\begin{align}
    \alpha_P = 
    \begin{cases}
        -\frac{n (\alpha_0 - 1)(P_s - P)^{n-1}}{(P_s - P_e)^{n}}, \qquad &\alpha=\alpha_{\rm max} \text{ and } \frac{dP}{dt} > 0 \qquad\text{ (compaction) }\\
        \alpha^2 \left(1 - \frac{1}{h^2}\right), \qquad & \alpha<\alpha_{\rm max} \text{ and/or } \frac{dP}{dt} < 0 \qquad\text{ (elastic) }
    \end{cases},
\end{align}
where
% \begin{align}
    $h = 1 + \tfrac{c_e - c_s}{c_s}\tfrac{\alpha - 1}{\alpha_0 - 1}.$
% \end{align}
%
Here, $c_e$ is the elastic sound speed for the porous material
and $c_s$ is the sound speed for the void-free material.
Therefore, the $P{\rm -}\alpha$ model for a porous material 
can be described using the parameters $(\rho_0, \rho_p, P_e, P_s, n, c_e, c_s)$.

% \begin{figure}[!htb]
%     \centering
%     \includegraphics[trim=5.5cm 0 1cm 0, clip, width=0.47\linewidth,  height=7cm,
%   keepaspectratio]{figures/initial.png}
%     \includegraphics[width=0.47\linewidth, height=7cm,
%   keepaspectratio]{figures/avgdens.png}
%     \caption{Left: initial configuration.
%     Right: evolution of average density.}
%     \label{fig:initial}
% \end{figure}

The material properties of the steel are assumed to be known
and given in \autoref{tab:steel_properties}.
Each of the material properties for the aluminum sample
are assumed to be unknown and bounded by defined ranges.
The ranges for the material parameters are given in \autoref{tab:al_properties}.
% In lieu of testing our approach on experimental data, we also consider a synthetic experiment where a model mismatch is introduced. In this experiment, the EoS and porosity model of
% the porous aluminum sample is replaced with a Sesame EoS model~\cite{sesameintro}.

\begin{table}[!htb]
\centering
\begin{tabular}{ccccccc}
\toprule
\multicolumn{5}{c}{Mie-Gr\"uneisen} & \multicolumn{2}{c}{Elastic Perfect Plasticity based on Von Mises yield surface} \\ \cmidrule(lr){1-5} \cmidrule(lr){6-7}
$\rho_0$ [$g/cm^3$] & $c_s$ [$cm/s$] & $s$ & $\Gamma_0$ & $C_V$ [$erg/(g\cdot eV)$] & Poisson's Ratio & Yield Strength [$dyne/cm^2$]\\
\midrule
7.81 & $4.58\cdot10^5$ & 1.49 & 1.69 & $5.091\cdot10^{10}$ & 0.3 & $9.45\cdot10^9$ \\
\bottomrule
\end{tabular}
\caption{Steel 4130 properties.}
\label{tab:steel_properties}
\end{table}

\begin{table}[!htb]
\centering
\begin{tabular}{ccccccccccc}
\toprule
& \multicolumn{5}{c}{Mie-Gr\"uneisen} & \multicolumn{5}{c}{$P{\rm -}\alpha$} \\
%\hline
\cmidrule(lr){2-6}\cmidrule(lr){7-11}
&$\rho_0$ & $c_s$  & $s$ & $\Gamma_0$ & $C_V$ & $\rho_p$ & $c_e$ & $P_s$  & $P_e$  & $n$ \\
\midrule
min & 2.707 & $5.2\cdot10^5$ & 1.32  & 2.24  & $1\cdot10^{11}$ & 2.1 & $1.5\cdot10^5$ & $4.5\cdot10^9$ & $4\cdot10^8$ & 1.9 \\
max & 2.815 & $5.25\cdot10^5$ & 1.38 & 2.48 & $1\cdot10^{11}$ & 2.175 & $5\cdot10^5$ & $1\cdot10^{10}$  & $8\cdot10^8$ & 2.2 \\ 
\midrule
units & [$g/cm^3$] & [$cm/s$] & -- & -- & [$erg/(g\cdot eV)$]& [$g/cm^3$] & [$cm/s$] & [$dyne/cm^2$] & [$dyne/cm^2$] & -- \\
\bottomrule
\end{tabular}
\caption{Aluminum 2024 property ranges.}
\label{tab:al_properties}
\end{table}

% The parameters of the $P{\rm -}\alpha$ model are often determined through experimental techniques:
% •	Crush-Curve Analysis: Experimental crush-curves, which depict the relationship between pressure and distention, are used to calibrate the $P{\rm -}\alpha$ model parameters for specific materials. 
% •	Shock Compression Experiments: Data from shock compression tests provide insights into the dynamic compaction behavior of porous materials, informing the development and validation of the $P{\rm -}\alpha$ model. 

% \section{Simulation of a flyer plate experiment}
\subsection{Data generation and machine learning models}
\label{sec:flyer}

\begin{figure}[!htb]
    \centering
    \includegraphics[width=0.9\textwidth]{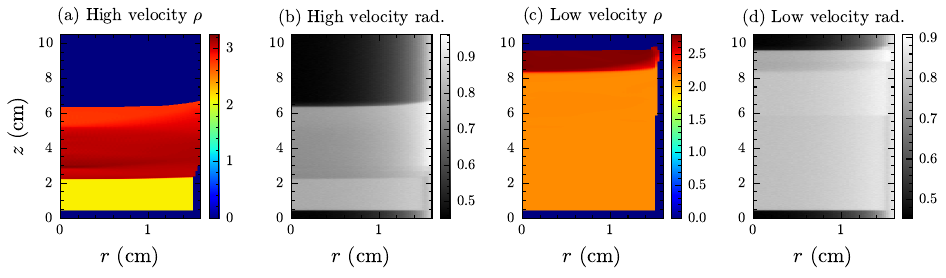}
    \caption{Visualizations of density field and noisy radiograph for one sample of EoS and $P-\alpha$ parameters, using high impact velocity ((a) and (b)) and low impact velocity ((c) and (d)).}
    % \caption{Visualizations of the density field and noisy radiograph for one test case. Left to right: (a) density field of 2024 aluminum, (b) noisy radiograph corresponding to the density field.}
    \label{fig:density_rad_feat_vis}
\end{figure}

The focus of this work is to use machine learning as a tool to identify data-spaces with well-posed mappings from observed density fields and radiographic images to EoS and $P-\alpha$ parameters, and to further construct machine learning models to infer such parameters from observed density fields or radiographic images. To generate data to analyze and train ML models, we sample an ensemble of 10,000 vectors of EoS and $P-\alpha$ parameters by randomly sampling each unknown aluminum material parameter independently and uniformly within its defined range (see \autoref{tab:al_properties}). We consider three choices for the initial impact velocity, $v_{\rm init}\in\{5\cdot10^4, 1\cdot10^5, 5\cdot10^5\}~\si{\cm/\second}$ (more on this in \autoref{sec:parameter_estimation}), and simulate the flyer plate experiment introduced previously for each velocity and each parameter vector (totaling 30,000 simulations) using the CTH simulation code~\cite{cth92}, which outputs a dynamic sequence of density profiles. In the $(r,z)$ domain we use a finite volume discretization on a structured quad mesh of $64 \times 488$ cells, and simulate for time $t\in[0,12\si{\us}]$ with adaptive explicit time integration. At the final simulated (observation) time of $t=12 \si{\us}$ we generate synthetic radiographic images from the resulting density fields using an imaging model analogous to the one described in~\cite{Serino24}. This model simulates contamination from several realistic sources of radiographic noise, including correlated blur, scatter, and a Poisson noise field. We provide complete specifications of the imaging geometry and noise model in \autoref{app:radiography}.
% This model involves first obtaining the areal mass
% for the steel and aluminum using the cone beam projection
% provided by the LEAP\cite{?}.
\autoref{fig:density_rad_feat_vis} shows example density fields at $t=12\si{\us}$ of the 2024 aluminum and the corresponding radiographic measurements for our low ($v_{\rm init}=5\cdot10^4~\si{cm/s}$) and high ($v_{\rm init}=5\cdot10^5~\si{cm/s}$) initial impact velocity of the steel flyer plate. \autoref{tab:figure_params} in \autoref{app:additional_figs} provides the EoS and $P{\rm -}\alpha$ parameters of the aluminum shown in this example, as well as figures through the rest of the manuscript to aid reproducibility.

To construct mappings from a space of observed density fields or radiographic images back to EoS and $P-\alpha$ parameters, we use a conditional variational autoencoder\cite{VAE, ssldgm} (VAE) machine learning architecture. Given radiographs or density fields, the proposed network architecture can be trained to sample from the corresponding posterior distribution of parameter values. When the network is trained to infer parameters from density fields, we refer to it as the density-to-parameters variational autoencoder (D2P-VAE), whereas when radiographs are used as the input, we refer to it as the radiographs-to-parameters variational autoencoder (R2P-VAE). The simulated data is split 80/10/10 into training, testing, and validation datasets, and the architecture and training details for the VAEs are described in further detail in \autoref{sec:AppR2PVAE}. We note here that we adopt a modern protocol for training VAEs known as the $\sigma$-VAE\cite{sigmavae}, which is both statistically principled and requires less manual hyperparameter tuning than similar VAE formulations. The practical effectiveness of this method may be of interest to other practitioners. Density reconstructions are achieved by combining parameter estimation with traditional physics simulation. Given noisy radiographs, we use the R2P-VAE to infer the relevant physical parameters, and these inferred parameters are then used as input to the CTH code to simulate the flyer plate impact and recover the evolution of the density field. For reference, the proposed approach of combining the R2P-VAE with forward simulation to reconstruct density fields from noisy radiographs is compared to the more common approach of using a neural network that directly reconstructs density fields from noisy radiographs in \autoref{app:R2D-Net}.
% \footnote{The proposed approach combining R2P-VAE with forward simulation to reconstruct density fields from noisy radiographs is compared to the more common approach of using a neural network that directly reconstructs density fields from noisy radiographs in \autoref{app:R2D-Net}.}

% We compare two approaches to density reconstruction. The first approach is to use the parameters predicted by R2P-VAE. Once the relevant physical parameters are predicted, the CTH code can be used to simulate the flyer plate impact and recover the evolution of the density field.

% The second approach, which we consider as a baseline, is to use a neural network to directly reconstruct density fields given noisy radiographs. The proposed radiographs-to-density network (R2D-Net) takes a pair of radiographs captured from low and high impact velocity experiments as input and directly produces the density fields for both experiments. The architecture we use is a U-Net\cite{unet} very similar to the networks used in\cite{deblurring_via_sr,bell2024supervised}. For full architecture and training details, we refer to \autoref{app:R2D-Net}.

% ---------------------------------------------------- %
% ---------------------------------------------------- %
% ---------------------------------------------------- %
\section{Parameter estimation results}
\label{sec:parameter_estimation}

% {We now present the results of parameter estimation and density reconstruction with the D2P-VAE and R2P-VAE. We first examine parameter estimation from the true density fields of the aluminum in \autoref{subsec:params_from_density}. Although these density fields are not directly observable experimentally, these studies are valuable because they reveal the information content of the experiments conducted at the three impact velocities. We show that observations from any single experiment are insufficient to enable accurate prediction of all of the parameters, whereas all of the parameters can be recovered by combining observations from high impact velocity and low impact velocity experiments. This finding informed the design of our subsequent studies on parameter estimation from noisy radiographs.} 

\subsection{Well-posed mappings from density fields to EoS and $P-\alpha$ parameters}
\label{subsec:params_from_density}

\begin{figure}[!htb]
    \centering
    \includegraphics[width=\linewidth]{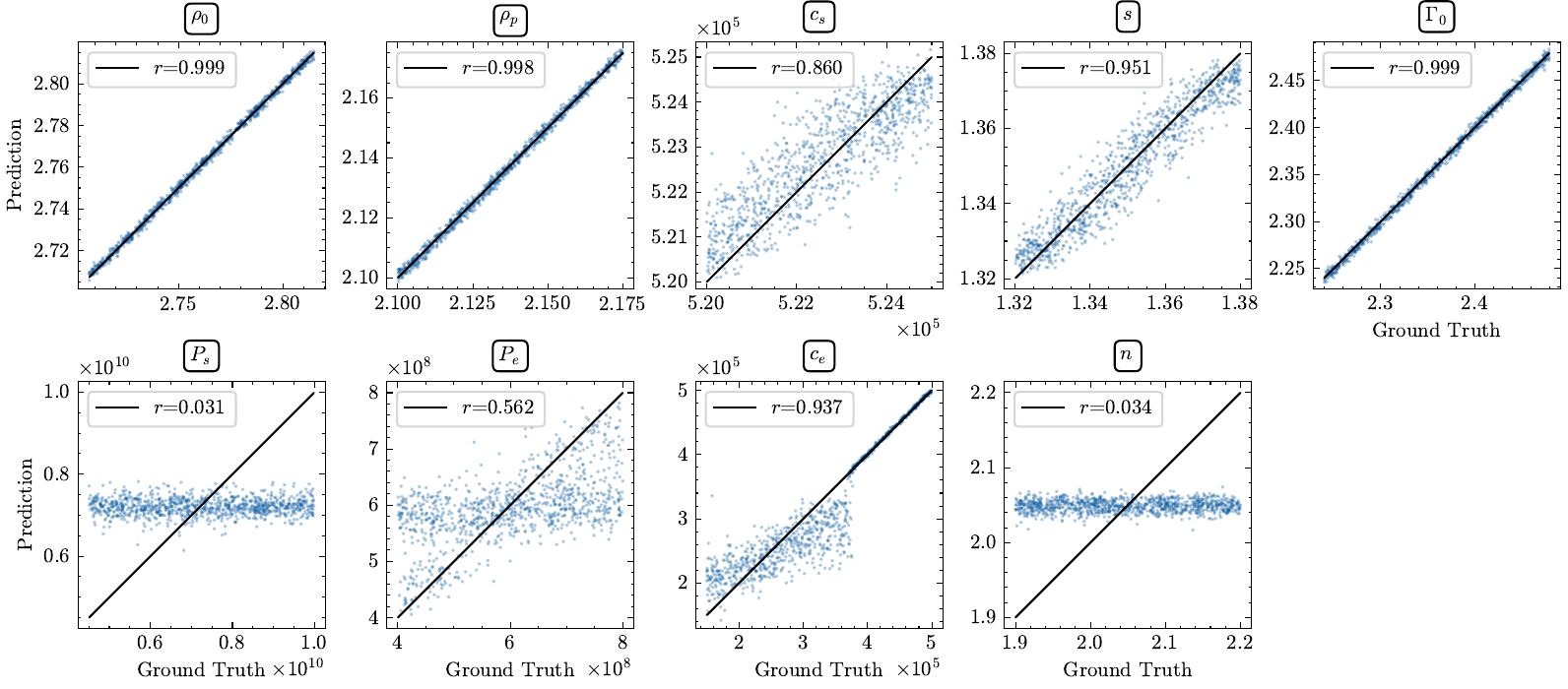}
    \caption{Parameter estimates from the trained D2P-VAE on the testing set when the density of the 2024 aluminum from the high impact velocity experiment is used as input. For each test case, we use the mean of 1,000 posterior samples as a point estimate of the parameter values. Identity lines are included for reference. We also report the Pearson correlation coefficient $r$ in each subplot.}
    \label{fig:high_vel_to_density}
\end{figure}

We begin by using ML with our generated flyer plate density data to explore the space of hydrodynamic density data with sufficient ``information content'' to establish a well-posed mapping to the underlying EoS and $P-\alpha$ parameters of interest.% to be established.
To motivate this examination, we first demonstrate likely degeneracy in mapping from a final observed density field to the physical parameters if only a single high impact velocity is considered. 
We trained the D2P-VAE to estimate parameters using the density field of the 2024 aluminum from the high impact velocity experiment ($v_{\rm init}=5\cdot10^5~\si{cm/s})$ at 12 $\si{\us}$. For every test case, the trained D2P-VAE was used to produce a posterior distribution of physical parameters given the density field, and these posterior distributions were used to produce point estimates of the parameter values for each test case. We calculated the mean of the posterior distribution to obtain the minimum mean squared error (MMSE) estimate of the parameter values for each test case. Assuming that the D2P-VAE samples from the true posterior of parameter values for a given density field, these point estimates will be optimal in terms of MSE and $r^2$ when compared with the ground truth parameter values. These point estimates are shown in \autoref{fig:high_vel_to_density}.

We observe that for this initial impact velocity, the network is able to recover the parameters $\rho_0$, $\rho_p$, and $\Gamma_0$ with high accuracy, and Mie-Gr\"uneisen parameters $c_s$ and $s$ with reasonable accuracy. However, there is almost no inference power for crush model parameters $P_s$, $P_e$, and $n$. Interestingly, we find that the elastic sound speed $c_e$ can determined very accurately if its true value is above approximately $3.8\cdot10^5~\si{cm/s}$, but is inferred with much less certainty otherwise. Due to the large quantity of training data and use of state-of-the-art ML architectures, these results provide high confidence that a well-posed mapping from a single observed density field to EoS and $P-\alpha$ parameters cannot be constructed for this impact velocity. We point out that this also makes sense physically--in \autoref{fig:density_rad_feat_vis}(a) we see that at 12 \si{\us} there is a large region of aluminum with density greater than $\rho_0$, while the remainder of the aluminum has density approximately equal to $\rho_p$. Thus our observable data consists entirely of fully compacted material and nearly uncompressed material, which does not provide sufficient information to completely constrain the multi-parameter crush model.
% but is providing insufficient physical information to constrain a multiparameter compaction model. 

% {We show the correlations between these MMSE parameter estimates and the true parameter values across the testing set in \autoref{fig:high_vel_to_density}. We find that This result shows that a single observation of the system is insufficient to enable recovery of all of the parameters of interest, even without the presence of noise in the observations. This result is not surprising, given the densities and pressures produced in the high impact velocity experiment. 

To facilitate improved experimental design and generation of data that can be used to properly resolve and disentangle EoS and $P-\alpha$ parameters, we now consider practical variations in the choice of observed density states to improve inference power. We consider (i) using a time-sequence of density profiles, corresponding to rapid successive imaging of the impact and compaction process, and (ii) using different initial flyer plate impact velocities (and some combinations thereof). For (i), we consider a training the D2P-VAE using a time series of density fields at times $t\in\{0, 2, 4, 6, 8, 10, 12\}\si{\us}$ from the high impact velocity experiment. For (ii) we consider three different initial velocities as specified previously, $v_{\rm init}\in\{5\cdot10^4, 1\cdot10^5, 5\cdot10^5\}~\si{\cm/\second}$, which we refer to as low, medium, and high impact velocities, respectively, and also consider combined training data using the high and medium impact velocities or high and low impact velocities. We trained the D2P-VAE architecture on each of these distinct datasets. In \autoref{tab:ccs_from_density}, we report the correlation coefficients of the MMSE predictions with the true parameter values across the testing set, as well as the mean absolute percentage error (MAPE) of the predictions, where the MAPE for a set of $N$ predictions $\hat{y}_i$ with corresponding ground truth values $y_i$ is calculated as $100\cdot\frac{1}{N}\sum_{i=1}^N \left| \frac{\hat{y}_i-y_i}{y_i} \right|$. 

\begin{table}[!htb]
    \centering
    % \large Parameter Estimate \\
    \begin{tabular}{ccccccccccc}
    \toprule
    % & & & \multicolumn{3}{c}{Mie-Gr\"uneisen} & \multicolumn{4}{c}{$P{\rm -}\alpha$} \\ \cmidrule(lr){4-6} \cmidrule(lr){7-10}
    Metric & Network Input & $\rho_0$ & $\rho_p$  & $c_s$ & $s$ & $\Gamma_0$ & $P_s$ & $P_e$  & $c_e$  & $n$ \\ \cmidrule(lr){1-1} \cmidrule(lr){2-2} \cmidrule(lr){3-11}
    \multirow{6}{*}{\shortstack[c]{Corr. coeff.\\(higher is\\better)}}
    & Low impact vel. & 0.997 & 0.996 & \color{red}0.843 & \color{red}-0.011 & 0.995 & 1.000 & 1.000 & 1.000 & 0.988 \\
    & Medium impact vel. & 0.996 & 0.995 & \color{red}0.025 & \color{red}0.029 & 0.971 & 0.987 & 0.989 & 1.000 & \color{red}0.230 \\
    & High impact vel. & 0.999 & 0.998 & \color{red}0.860 & 0.951 & 0.999 & \color{red}0.031 & \color{red}0.562 & 0.937 & \color{red}0.034 \\
    & High impact vel. time series & 0.999 & 0.999 & 0.980 & 0.998 & 1.000 & \color{red}0.051 & \color{red}0.484 & 0.971 & \color{red}0.021 \\
    & High and medium impact vel. &0.999 & 0.998 & \color{red}0.883 & 0.936 & 0.998 & 0.988 & 0.982 & 1.000 & \color{red} \color{red}0.152 \\
    & \bf High and low impact vel. & 0.999 & 0.998 & 0.922 & 0.953 & 0.999 & 1.000 & 0.998 & 1.000 & 0.977 \\
    \midrule
    
    \multirow{6}{*}{\shortstack[c]{MAPE\\(lower is\\better)}}
    & Low impact vel. & 0.078 & 0.084 & 0.114 & \color{red}1.138 & 0.241 & 0.332 & 0.408 & 0.297 & 0.457 \\
    & Medium impact vel. & 0.074 & 0.080 & 0.239 & \color{red}1.086 & 0.571 & \color{red}2.841 & \color{red}1.256 & 0.276 & \color{red}3.638 \\
    & High impact vel. & 0.042 & 0.046 & 0.115 & 0.318 & 0.114 & \color{red}20.19 & \color{red}13.67 & \color{red}9.175 & \color{red}3.751 \\
    & High impact vel. time series & 0.032 & 0.042 & 0.045 & 0.059 & 0.048 & \color{red}20.77 & \color{red}14.73 & \color{red}6.551 & \color{red}3.767 \\
    & High and medium impact vel. & 0.044 & 0.045 & 0.105 & 0.351 & 0.136 &\color{red} 2.767 & \color{red}1.562 & 0.343 & \color{red}3.684 \\
    & \bf High and low impact vel. & 0.048 & 0.050 & 0.086 & 0.305 & 0.113 & 0.604 & 0.924 & 0.505 & 0.814 \\
    \bottomrule
    \end{tabular}
    \caption{Pearson correlation coefficients and mean absolute percentage errors (MAPE) between the parameters inferred by the D2P-VAE and the ground truth values across the testing set for different inputs. To compute both metrics, we used the mean of the inferred posterior as a point estimate of the parameter values for each test case.
    %The inputs considered are the density fields of the aluminum at $12~\si{\us}$ from impacts with different values of $v_{\rm{init}}$ (except "high impact vel. time series," which uses observations at 0, 2, 4, 6, 8, 10, and 12~$\si{\us}$). The low, medium, and high impact velocities are $5\cdot10^4$, $1\cdot10^5$, and $5\cdot10^5~\si{\cm/\second}$.
    Correlation coefficients $<0.9$ and MAPE $>1\%$ are marked in {\color{red}red}.}
    \label{tab:ccs_from_density}
\end{table}

\begin{figure}[!htb]
    \centering
    \includegraphics[width=\linewidth]{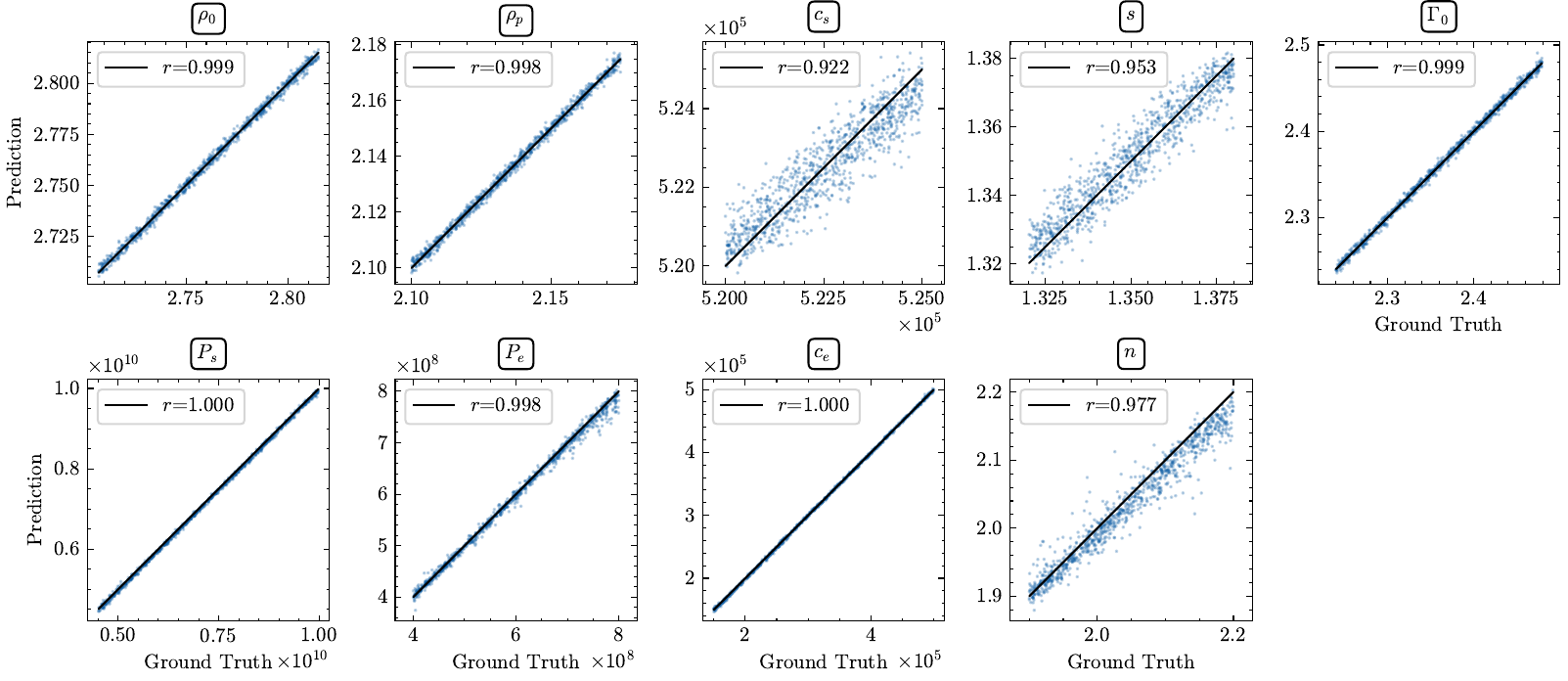}
    \caption{Parameter estimates from the trained D2P-VAE on the testing set when the densities of the 2024 aluminum from the high impact velocity experiment and the low impact velocity experiment are used as input. For each test case, we use the mean of 1,000 posterior samples as a point estimate of the parameter values. Identity lines are included for reference. We also report the Pearson correlation coefficient $r$ in each subplot.}
    \label{fig:high_vel_low_vel_to_density}
\end{figure}

The results in \autoref{tab:ccs_from_density} indicate that using additional temporal observations with a high impact velocity in approach (i) slightly improves the inference of the Mie-Gr\"uneisen parameters $c_s$ and $s$, but yields effectively no meaningful improvement in inference of $P-\alpha$ model parameters $P_s$, $P_e$, and $n$. This is potentially because all of the observed density fields only contain regions of fully compacted aluminum (see the time series in  \autoref{fig:initial}, for example), and hence do not sufficiently span the potential compaction dynamics and behaviors to resolve the underlying parameters. In contrast, using only a low impact velocity yields excellent inference of $P-\alpha$ parameters, but poor inference of $s$ and suboptimal prediction of $c_s$. The medium impact velocity maintains some of the benefit of the low impact velocity, but is not able to resolve the $P-\alpha$ parameter $n$. Moreover, the EoS parameters $c_s$ and $s$ cannot be inferred as accurately as when the high impact velocity is used.
Our intuition is that we need to shock the material (with a high impact velocity) to effectively recover the EoS parameters, while the material needs to be crushed more slowly to accurately infer the porosity parameters.
% Note that this is in contrast to what we believe is needed to accurately infer porosity parameters.
Following this intuition, we see that by combining high and low impact velocity data for training the D2P-VAE, we are able to maintain the inference power of each set of data (that is, accuracy with respect to EoS and $P-\alpha$, respectively), and this combination is the only scenario we have tested that enables good inference power in both (or either) correlation coefficient and MAPE metrics. The parameters predicted by the D2P-VAE trained with combined low and high impact velocity data across the testing set are shown in \autoref{fig:high_vel_low_vel_to_density}.

%We further examine the densities produced in the different experiments in \autoref{fig:max_density}, where we show the maximum density achieved over the entire 12~$\si{\us}$ window versus the initial flyer velocity, as well as the maximum velocity over time for different flyer velocities. We find that %the maximum density %produced in the high %impact velocity experiment %is substantially high than %in the low or medium %impact velocity %experiments. %Moreover, %the highest density is %produced almost %immediately, within about %0.2 $\si{\us}$ from the %time of impact.

%\begin{figure}
%    \centering
    %\includegraphics[width=0.8\linewidth]{Flyer_Plate/DensityVsVelocityBoth.pdf}
 %   \caption{Maximum densities observed during flyer plate impact experiments. Subfigure (a) shows the maximum density of the aluminum over the entire 12 $\si{\us}$ simulation for flyer velocities between $5~\si{cm/s}$ and $5\cdot10^5~\si{cm/s}$, while subfigure (b) shows the maximum density of the aluminum over time for five different impact velocities, including the low, medium, and high impact velocities used in our experiments (orange, red, and purple lines).}
    \label{fig:max_density}

\subsection{Parameter inference from (noisy) radiographs}

In the preceding section we determined combined density observables that allow for robust inference of EoS and $P-\alpha$ parameters. However, in the experimental setting we are typically not able to obtain direct high-fidelity measurements of density, and instead we must work with radiographic projections. Following the experimental design determined in the previous section, we train the R2P-VAE architecture to infer parameters using synthetic radiographs from combined high and low impact velocity data. We first examine using clean radiographs as input to the network, followed by noisy radiographs to better understand the impact of progressive loss of information content in the observations.
%from which parameter inference is determined.
For every test case, the trained R2P-VAE is used to produce a posterior distribution of physical parameters given the pair of radiographs, and MMSE parameter estimates are computed by taking the mean of the posterior for each case. Correlation coefficients and MAPE with respect to ground truth using clean density fields, clean radiographs, and noisy radiographs are shown in \autoref{tab:ccs_comparison}. Plots of parameter inference across the full testing set for clean and noisy radiographs are provided in \autoref{app:CsAndS} in \autoref{fig:clean_rads_correlations} and \autoref{fig:MMSE_param_estimates_new}, respectively. 
    
\begin{table}[!htb]
    \centering
    % \large Parameter Estimate \\
    \begin{tabular}{ccccccccccc}
    \toprule
    % & & & \multicolumn{3}{c}{Mie-Gr\"uneisen} & \multicolumn{4}{c}{$P{\rm -}\alpha$} \\ \cmidrule(lr){4-6} \cmidrule(lr){7-10}
    Metric & Observable & $\rho_0$ & $\rho_p$  & $c_s$ & $s$ & $\Gamma_0$ & $P_s$ & $P_e$  & $c_e$  & $n$ \\ \cmidrule(lr){1-1} \cmidrule(lr){2-2} \cmidrule(lr){3-11}
    \multirow{3}{*}{\shortstack[c]{Corr. coeff.\\(higher is\\better)}}
    & Density field & 0.999 & 0.998 & 0.922 & 0.953 & 0.999 & 1.000 & 0.998 & 1.000 & 0.977 \\
    & Clean radiograph & 1.000 & 1.000 & {\color{red}0.791} & 0.906 & 0.997 & 0.999 & 0.984 & 1.000 & 0.927 \\
    & Noisy radiograph & 0.997 & 0.998 & {\color{red}0.473}&{\color{red}0.474} & 0.988 & 0.983 & {\color{red}0.858} & 1.000 & {\color{red}0.188} \\
    \midrule
    
    \multirow{3}{*}{\shortstack[c]{MAPE\\(lower is\\better)}}
    & Density field & 0.048 & 0.050 & 0.086 & 0.305 & 0.113 & 0.604 & 0.924 & 0.505 & 0.814 \\
    & Clean radiograph & 0.022 & 0.016 & 0.135 & 0.420 & 0.167 & 0.826 & {\color{red}2.704} & 0.497 & {\color{red}1.213} \\
    & Noisy radiograph & 0.066 & 0.054 & 0.206 & 0.951 & 0.373 & {\color{red}3.488} & {\color{red}8.482} & 0.567 & {\color{red}3.658} \\
    \bottomrule
    \end{tabular}
    \caption{Pearson correlation coefficients and mean absolute percentage errors (MAPE) of parameters inferred over the testing set by the VAE architecture trained on full density fields, clean radiographs, and noisy radiographs. To compute both metrics, we used the mean of the inferred posterior as a point estimate of the parameter values for each test case. Correlation coefficients $<0.9$ and MAPE $>1\%$ are marked in {\color{red}red}.}
    \label{tab:ccs_comparison}
\end{table}

\begin{figure}[!htb]
    \centering
    \includegraphics[width=0.65\linewidth]{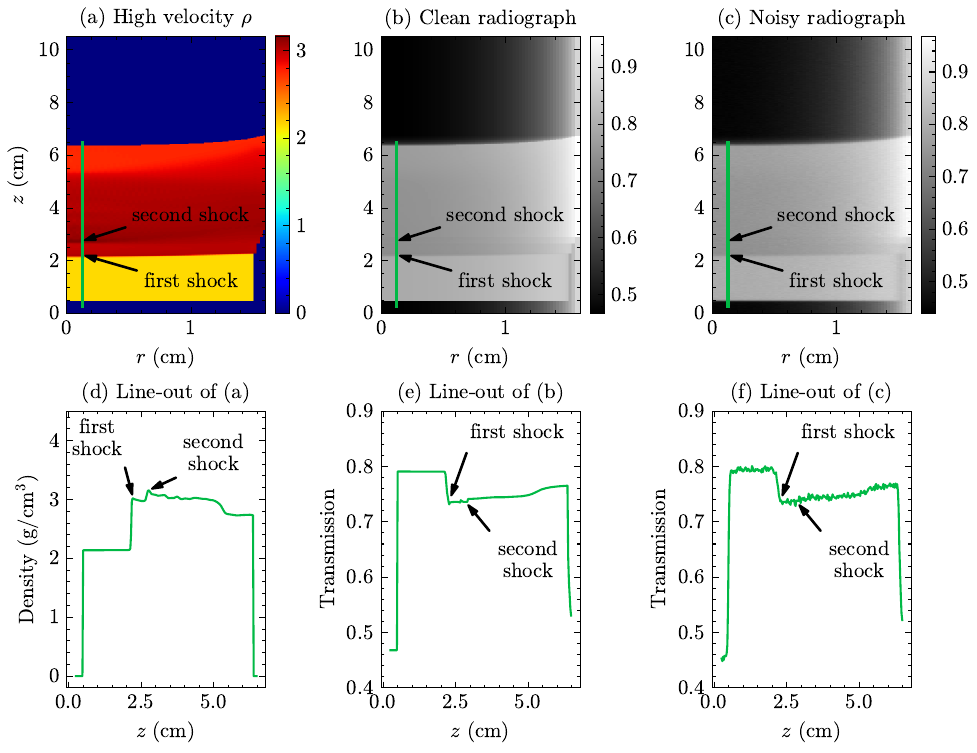}
    \caption{Line-outs of the density field, clean radiograph, and noisy radiograph for the high impact velocity experiment of one test case. The green vertical line located at $r=0.125$cm in subfigures (a), (b), and (c) indicates the location of the line-outs shown in subfigures (d), (e), and (f). In all subfigures we have annotated the positions of the first and second shocks, which can be seen most clearly in the density field.}
    \label{fig:line_outs}
\end{figure}

Note that \autoref{tab:ccs_comparison} demonstrates the distinction between correlation coefficient and MAPE as inference accuracy metrics. Although MAPE provides a direct measure of relative error of inferred parameter value, MAPE appears artificially low for parameters that only span a small relative range, e.g., $c_s$ and $s$ with noisy radiographs have poor inference (this can be observed visually \autoref{fig:MMSE_param_estimates_new} in \autoref{app:CsAndS}) but low MAPE. For clean radiographs we have only a modest loss in inference accuracy, but are still able to obtain overall good correlation and low MAPE for all parameters (this can be observed visually in \autoref{fig:clean_rads_correlations} in \autoref{app:CsAndS}). In contrast, using noisy radiographs we lose significant inference power in $c_s$, $s$, and $n$. This indicates the inability to recover these parameters largely results from the noise and scatter in the radiographic forward model. To understand this loss of inference power, we visualize the primary shock features found in the simulated density fields, clean radiographs, and noisy radiographs for the high impact velocity setting in \autoref{fig:line_outs}, including line-outs with prominent physical features annotated. In the density field, both the first shock, which has the largest density jump, and the second shock, which has the greatest density, are clearly visible. In the clean radiograph, which is formed by applying an Abel transform to the density field, the first shock is still quite visible, but the relative magnitude of the second shock is greatly diminished. After the addition of the radiographic noise, however, the second shock is completely obscured, and only the first shock can be located robustly.

\begin{figure}[!htb]
    \centering
    \includegraphics[width=0.8\linewidth]{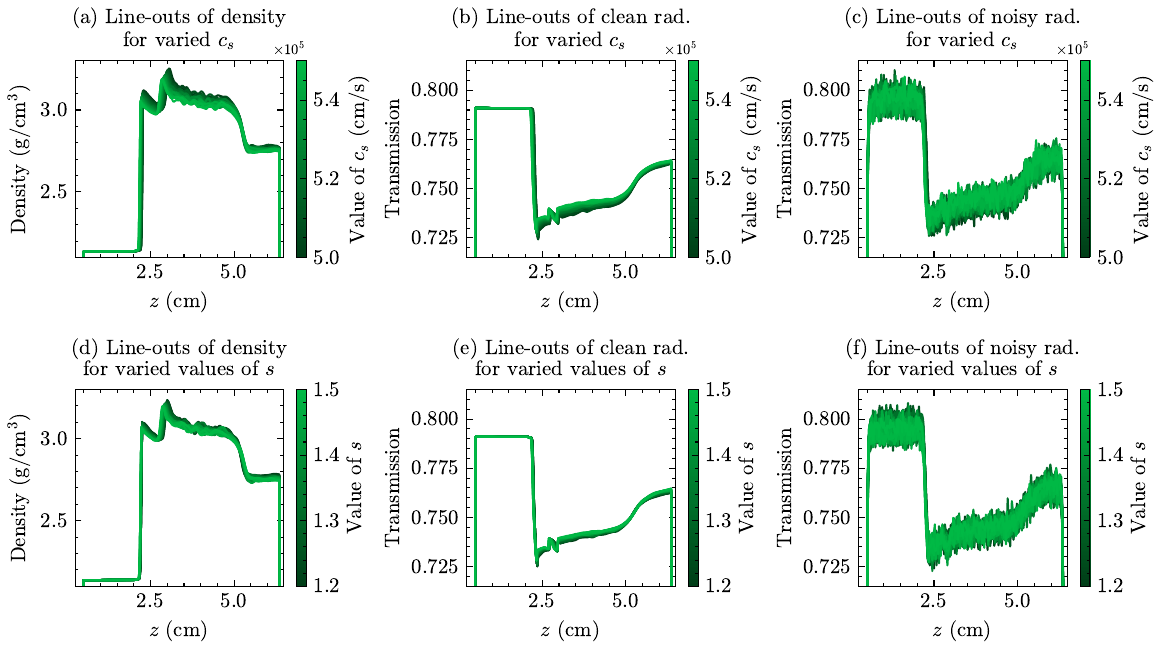}
    \caption{Line-outs of the density field, clean radiograph, and noisy radiograph for high impact velocity experiments generated with ranges of $c_s$ and $s$ values. Note that the ranges of the parameter values shown here are substantially exaggerated compared to the ranges used elsewhere (compare with \autoref{tab:al_properties}), and were chosen to enhance visibility of the variations.}
    \label{fig:line_outs_varied_cs_s}
\end{figure}

Sticking to the high impact velocity case as in \autoref{fig:line_outs}, we consider the sensitivity of solution line-outs from \autoref{fig:line_outs} to variation in $c_s$ and $s$ (because these parameters are robustly determined in the high impact velocity setting) in \autoref{fig:line_outs_varied_cs_s}. In each subplot, only one parameter value varies, while all others are fixed. Despite considering parameter variations much larger than what is used in the rest of this work (see \autoref{tab:al_properties}), we see that these significant variations in $c_s$ and $s$ only yield small variations in the line-outs of the integrated solution. The reduced relative magnitude of variation in clean radiographs leads to the modest reduction in inference power, but we see that solution sensitivity to parameter variation in $c_s$ and $s$ is effectively lost in the noisy radiographs, explaining the relatively poor inference of these parameters in \autoref{tab:ccs_comparison}.

\begin{figure}[!htb]
    \centering
    \includegraphics[width=0.9\linewidth]{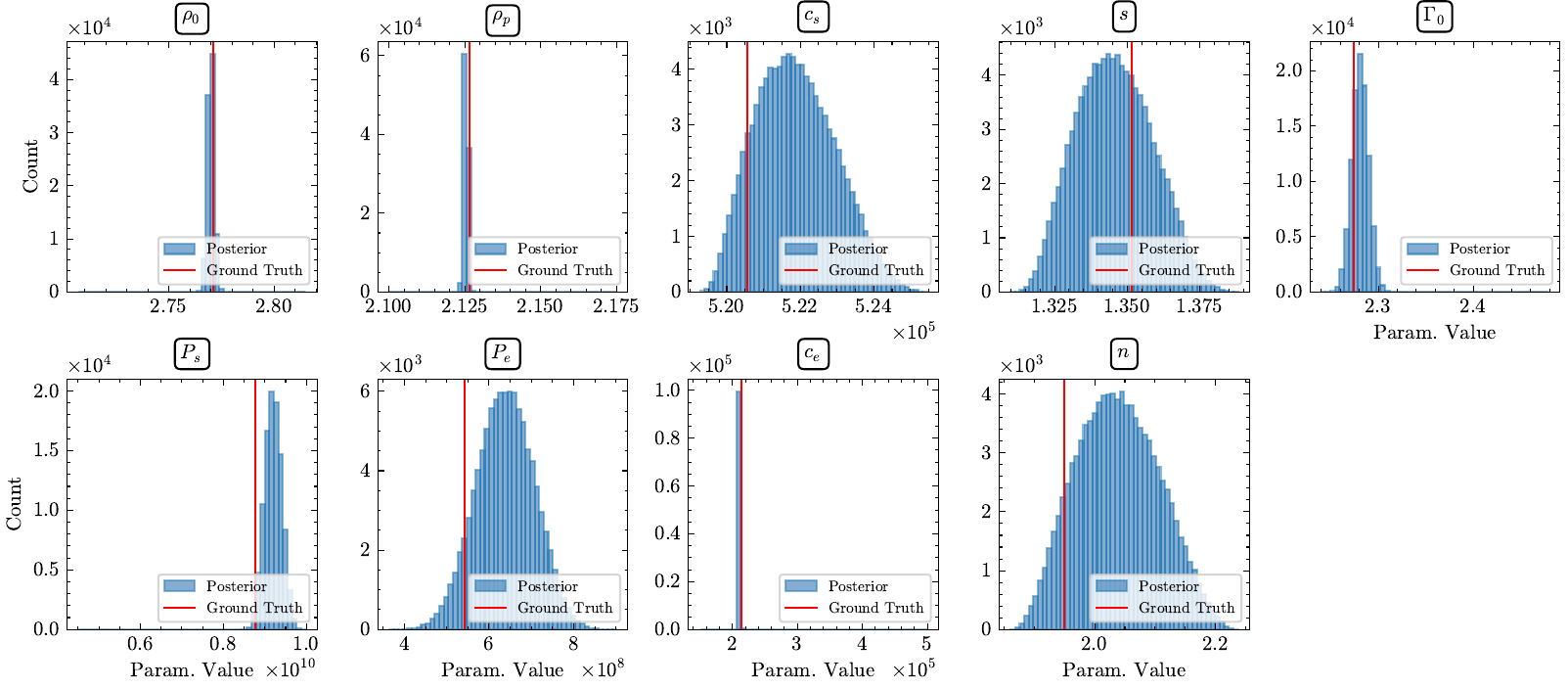}
    \caption{Representative posterior distribution predicted by the trained R2P-VAE on the testing set. We show the distribution of 100,000 posterior samples for the test case. Histograms show the distributions of predicted parameter values and vertical lines are located at the true parameter values for the test case.}
    \label{fig:posteriors_new}
\end{figure}

A benefit of the generative nature of R2P-VAE is that its output distribution automatically provides a proxy for confidence/uncertainty in the parameter estimates. \autoref{fig:posteriors_new} shows the inferred posterior distribution from the noisy radiographs for a representative test case, along with the ground truth parameter values. The chosen test case has the median error across the test set, where ``error'' is based on the $L^2$ distance between the posterior mean and the nominal parameter values, computed after $z$-score normalization of parameter values. Referring back to \autoref{tab:ccs_comparison}, the width of inferred distributions in \autoref{fig:posteriors_new} corresponds well with accuracy of the trained network applied to the test set, with parameters $n$, $s$, $c_s$, and $P_e$ having the respectively broadest inferred distributions as well as the lowest correlation coefficients. However, we note that the ground truth parameter value always lies within the inferred posterior in \autoref{fig:posteriors_new}.

\begin{figure}[!htb]
    \centering
    \includegraphics[width=0.9\linewidth]{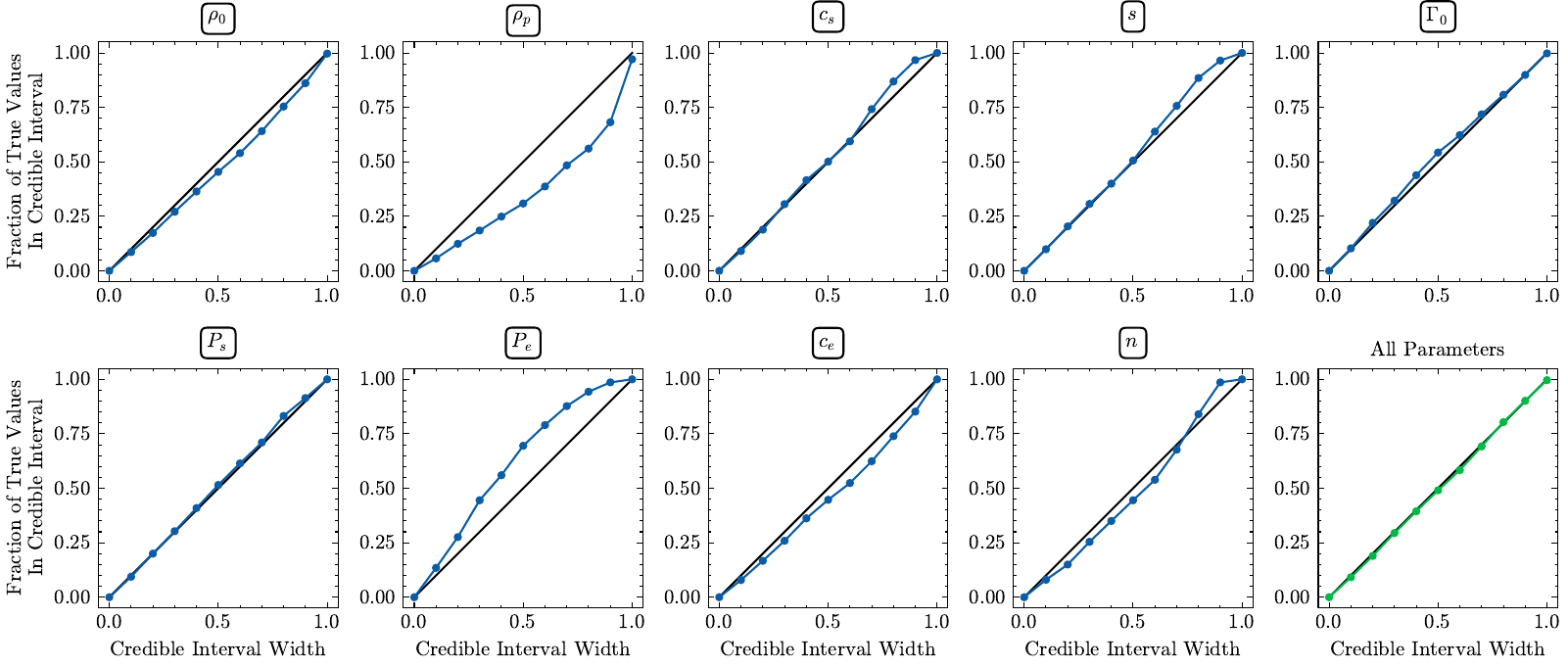}
    \caption{Posterior calibration or empirical coverage plots for the posteriors predicted by the R2P-VAE. For every test case, we used the predicted posterior to calculate credible intervals of various widths for each marginal distribution of parameter values. The points in each plot indicate the proportion of nominal values that actually lie in the credible intervals of a given width. The lower right plot (green line) shows the overall calibration, which is computed using all parameters. The remaining plots show the calibration for the marginal distributions of each parameter. Points near the identity line indicate well-calibrated uncertainty estimates, while points below the identity line represent overconfident predictions, and points above the line indicate underconfidence.}
    \label{fig:calibration}
\end{figure}

To further quantify the utility of the R2P-VAE in producing uncertainty estimates, in \autoref{fig:calibration} we provide a posterior calibration or empirical coverage plot, which shows how well credible intervals estimated from the predicted posteriors actually cover the nominal parameter values. For each test case, we computed quantile-based credible intervals for the marginal distribution of each parameter using samples drawn from the R2P-VAE. We then computed the ``coverage'' of these intervals, i.e. the proportion of test cases for which the nominal value lies in the credible interval. \autoref{fig:calibration} shows the empirical coverage for various credible interval widths. We find that the overall coverage is very well calibrated. The marginal coverage for each parameter is generally accurate, except the R2P-VAE appears somewhat overconfident in predicting $\rho_p$ and underconfident in predicting $P_e$. For example, the 50\% credible intervals for $\rho_p$ only contain the true value for about 30\% of cases, while the 50\% credible intervals for $P_e$ contain the true value for about 70\% of cases. We note, however, that computing credible \textit{intervals} from the marginal distributions of each parameter value provides limited information about the predicted credible \textit{region} in the full parameter space, so this apparent over- and underconfidence does not necessarily indicate a substantial miscalibration of the posterior.

\section{Density reconstruction results}
\label{sec:density_recon}

\begin{figure}[!htb]
    \centering
    \includegraphics[width=0.9\linewidth]{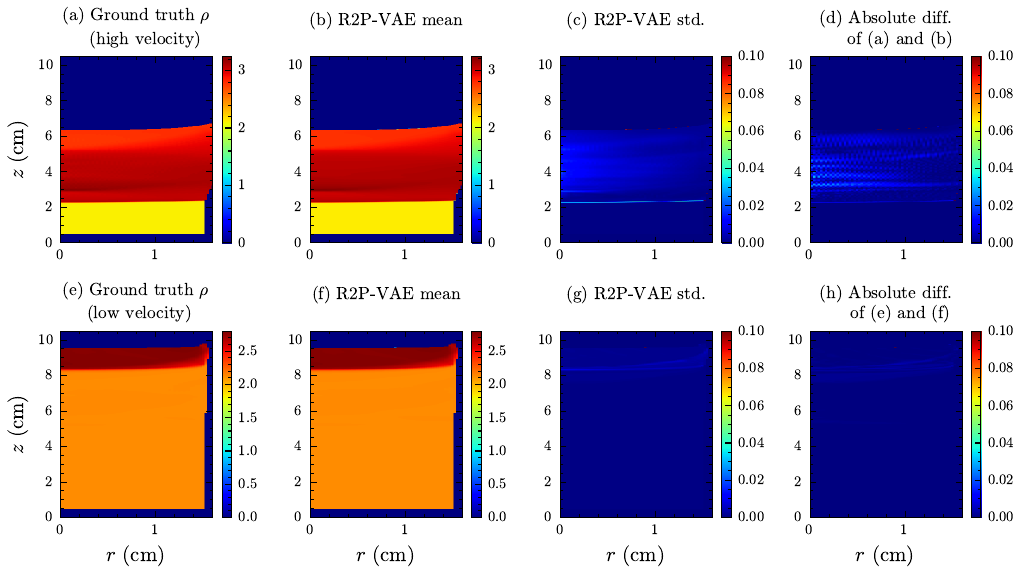}
    
    \caption{Density reconstructions and associated statistics using the parameters predicted by the R2P-VAE for a representative test case (the same test case as in \autoref{fig:posteriors_new}) and 1,000 posterior samples. Top row: high impact velocity, bottom row: low impact velocity. 
%    We ran 1,000 parameter samples from the R2P-VAE through the CTH code to produce 1,000 density fields at each impact velocity. We then computed the pixel-wise mean and standard deviation of the 1,000 density fields for both impact velocities. The top row of figures corresponds to the high impact velocity, and the bottom row corresponds to the low impact velocity. Subfigures: (a) ground truth density field, (b) mean of density reconstructions from 1,000 posterior samples, (c) standard deviation of density reconstructions from 1,000 posterior samples, (d) absolute difference of (a) and (b); subfigures (e-h) display the corresponding reconstructions at the lower impact velocity. Note the scale of the colorbars in (c), (d), (g), and (h), which have been truncated at 0.1 g/cm$^3$ for visibility.
}
    \label{fig:median_r2p_test_recon}
\end{figure}

{We produce density reconstructions from noisy radiographs for every case in the testing set by using the parameters inferred by the R2P-VAE as input to the CTH hydrocode.} In contrast to traditional approaches mapping directly from noisy radiographs to density fields, each instance of our approach yields a physically admissible density field that results from the nonlinear forward evolution of the physical equations. Representative average density reconstructions for a fixed set of EoS and $P-\alpha$ parameters are shown in \autoref{fig:median_r2p_test_recon} for the low and high impact velocity. We generated 1,000 samples from the posterior distribution of parameters using the R2P-VAE, and ran each of these parameter vectors through the CTH simulation code at the high and low impact velocities to produce 1,000 density fields for each impact velocity. We display the mean of the 1,000 reconstructions at each velocity, as well as the standard deviation of the reconstructions, and see that the R2P-VAE approach produces density reconstructions that are close to the ground truth. Note that the \emph{mean} density field is not necessarily physically admissible, but we display this primarily to demonstrate the high accuracy and low variance of density fields resulting from sampling the R2P-VAE network. As an additional point of reference, in \autoref{app:R2D-Net} we provide a comparison of the R2P-VAE approach to a radiographs-to-density network (R2D-Net) trained specifically for density reconstruction. Broadly, we find that the R2P-VAE provides comparably accurate reconstructions, with R2D-Net typically yielding lower RMSE while the R2P-VAE approach yields lower MAE.  However, we again emphasize that the R2P-VAE density reconstructions are consistent with both thermodynamic and the conservation laws while the R2D-Net does not adhere to physical laws.

%We find that The R2P-VAE method provides a strong prior on the solution, since the initial conditions of the experiment are known exactly.
%{We find that this prior generally prevents errors in the regions where there is no aluminum, while the largest uncertainties (and largest errors) are concentrated around the material interfaces.}

\begin{figure}[!htb]
    \centering
    \includegraphics[width=0.7\linewidth]{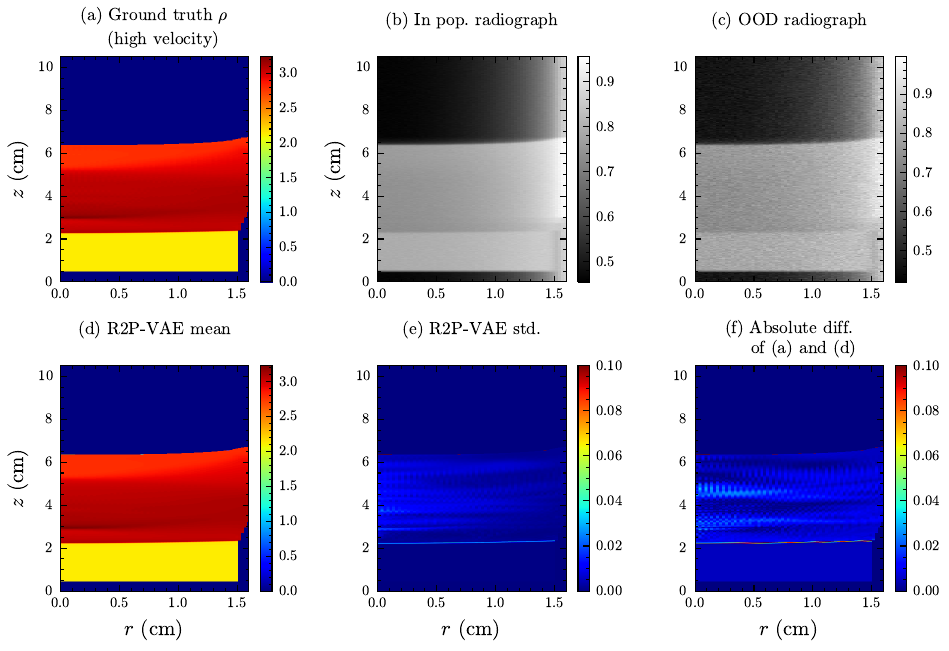}
    \caption{Demonstration of R2P-VAE density reconstruction pipeline for out of distribution (OOD) radiographic noise, using 1,000 posterior samples.}
    \label{fig:ood_noise}
\end{figure}

To test on scenarios that would arise in real experimental data, we next apply the R2P-VAE to radiographs that contain out-of-distribution (OOD) noise, i.e. noise models that the R2P-VAE has not observed in training. \autoref{fig:ood_noise} displays an in-distribution noisy radiograph for one test case, along with the same radiograph corrupted by OOD noise. The relative noise level is about five times greater for the OOD radiograph. {The remaining subplots in \autoref{fig:ood_noise} show the mean of 1,000 reconstructions from the R2P-VAE on this OOD data, the standard deviation of the reconstructions, and the error map. We again find that the density reconstructions are qualitatively accurate, and the largest errors occur along the material interfaces and the front shock.}

We quantify the performance of the R2P-VAE for density reconstruction in \autoref{tab:error_table_r2p}. For each of the 1,000 cases in the testing set, the R2P-VAE was used to draw 10 samples from the posterior distribution of parameters, each of which was run through the CTH code at high and low impact velocities to obtain density reconstructions. For each test case, error metrics were computed for each of the 10 reconstructions. These were then averaged across the entire set of reconstructions (10 reconstructions $\times$ 1,000 test cases) to obtain the mean errors listed under ``R2P-VAE samples'' in \autoref{tab:error_table_r2p}. Standard deviations of the errors for each test case were also computed. The mean of the computed standard deviations is reported after the $\pm$ in \autoref{tab:error_table_r2p} to give a sense of the variability in the errors. Point estimates of the reconstructed density were also computed by taking the mean of the 10 reconstructions for each test case. Error metrics and averaged results across the testing set for these reconstructions are presented under ``R2P-VAE mean.''
% We computed the metrics using the samples directly (R2P-VAE samples), as well as after taking the mean of 10 samples (R2P-VAE mean), which we believe may be closer to the mean of the posterior distribution of possible densities. 

The root mean squared error (RMSE), mean absolute error (MAE), and $L^\infty$ error averaged across the test set were computed for both the in-population data and OOD radiographic data, as displayed in \autoref{fig:ood_noise}. These metrics were also computed over a smaller region of interest (ROI) defined as $(r,z)\in[0\si{cm}, 1.5\si{cm}] \times [1.25\si{cm}, 6.0\si{cm}]$ for the high impact velocity and $(r,z)\in[0\si{cm}, 1.5\si{cm}] \times [7.75\si{cm}, 9.5\si{cm}]$ for the low impact velocity, to focus on error in the dynamic material and shock profiles, without considering material interfaces. For reference, a visualization of these regions is provided in \autoref{fig:roi}. Note that the $L^\infty$ error is significantly better in the ROI; this is because the large $L^\infty$ error results from $\mathcal{O}(1)$ cells at the material interface differing between the prediction and ground truth.
% . The regions of interest used are depicted in \autoref{fig:roi} in \autoref{app:additional_figs}.
%
We also find that we obtain accurate reconstructions in the presence of OOD noise, although the errors are generally about twice as large as for the in-population data. In general, these results demonstrate that our proposed method of combining parameter estimation and physical simulation can provide accurate density reconstructions, even when there are mismatches or unknowns in the forward model of the experiment.

\begin{table}[!ht]
    \centering
    \begin{tabular}{ccccc}
    \toprule
     &  & R2P-VAE mean & R2P-VAE samples \\ \midrule
    \multirow{2}{*}{RMSE} & In-Population & $1.98\cdot 10^{-2}$ & $(2.71\pm.60)\cdot 10^{-2}$   \\
     & OOD Noise & $3.37\cdot 10^{-2}$ & $(3.93\pm.55)\cdot 10^{-2}$   \\ \midrule
     
     \multirow{2}{*}{MAE} & In-Population & ${1.84 \cdot 10^{-3}}$ & $(2.20\pm.37)\cdot 10^{-3}$   \\
     & OOD Noise & ${4.42\cdot 10^{-3}}$ & $(4.66\pm.48)\cdot 10^{-3}$  \\ \midrule
     
     \multirow{2}{*}{Mean $L^\infty$ Error} & In-Population & $2.203$ & $2.818\pm.205$   \\
     & OOD Noise & $2.743$ & $2.944\pm.091$  \\ \midrule
     
     \multirow{2}{*}{RMSE over ROI} & In-Population & $5.96 \cdot 10^{-3}$ & $(7.91\pm.74)\cdot 10^{-3}$   \\
     & OOD Noise & ${1.36\cdot 10^{-2}}$ & $(1.46\pm.18)\cdot 10^{-2}$  \\ \midrule
     \multirow{2}{*}{MAE over ROI} & In-Population & ${3.64 \cdot 10^{-3}}$ & $(4.67\pm.63)\cdot 10^{-3}$   \\
     & OOD Noise & ${7.61\cdot 10^{-3}}$ & $(8.33\pm.73)\cdot 10^{-3}$ \\ \midrule
     \multirow{2}{*}{Mean $L^\infty$ Error over ROI} & In-Population & $4.64\cdot10^{-2}$ & $(6.34\pm1.74)\cdot 10^{-2}$   \\
     & OOD Noise & $1.30\cdot10^{-1}$ & $(1.38\pm.30)\cdot10^{-1}$ \\
     \bottomrule
    \end{tabular}
    \caption{Table of root mean squared error (RMSE) and mean absolute error (MAE) of density reconstruction methods for both in-population radiographic data and radiographs corrupted with out-of-distribtion (OOD) noise. We also report the metrics computed over a region of interest, which is $(r,z)\in[0\si{cm}, 1.5\si{cm}] \times [1.25\si{cm}, 6.0\si{cm}]$ for the high impact velocity and $(r,z)\in[0\si{cm}, 1.5\si{cm}] \times [7.75\si{cm}, 9.5\si{cm}]$ for the low impact velocity (visualization provided in \autoref{fig:roi}).}
    \label{tab:error_table_r2p}
\end{table}

\subsection{Physical model mismatch}

\begin{figure}[!htb]
    \centering
    \includegraphics[width=0.9\linewidth]{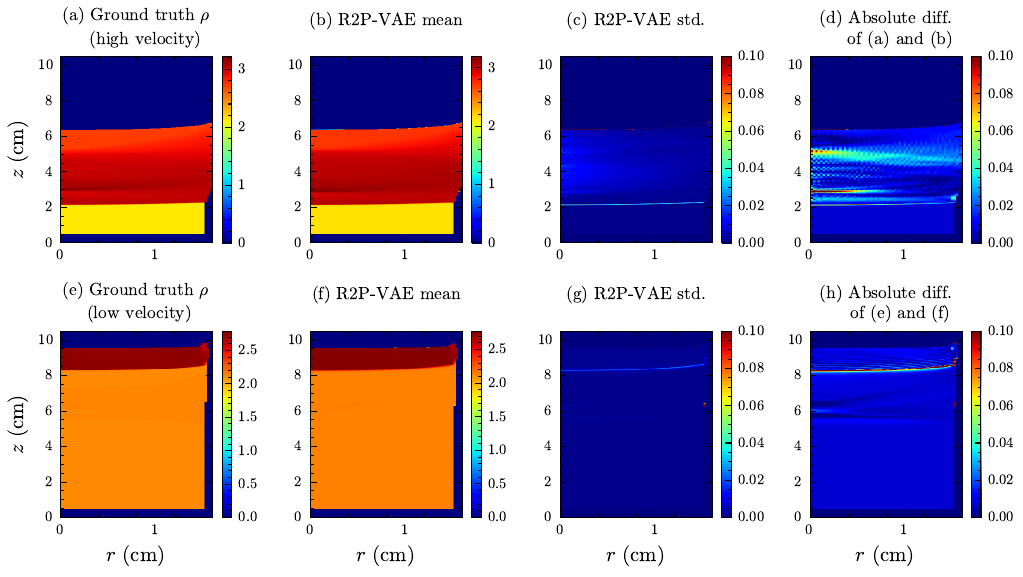}
    \caption{Density reconstructions using parameters predicted by the R2P-VAE for a test case that was generated using a mismatched EoS model. The R2P-VAE was trained using a Mie-Gr\"uneisen EoS, while the test case uses a Sesame EoS. 1,000 samples of parameters from the R2P-VAE were run through the CTH code to produce 1,000 reconstructions at each impact velocity. Subfigures: (a) ground truth density field, (b) mean of density reconstructions from 1,000 posterior samples, (c) standard deviation of density reconstructions from 1,000 posterior samples, (d) absolute difference of (a) and (b); subfigures (e-h) display the corresponding reconstructions at the lower impact velocity. Note the scale of the colorbars in (c), (d), (g), and (h), which have been truncated at 0.1 g/cm$^3$ for visibility.}
    \label{fig:sesame_visualization}
\end{figure}

To further demonstrate the potential application of the proposed method to experimental data, we apply the R2P-VAE to data where the underlying EoS parameterizations are different than those used for training. In particular, we generated a pair of high and low impact velocity experiments where the EoS is replaced with a Sesame EoS model~\cite{sesameintro}.
We then generated noisy radiographs for these experiments and passed them through the trained R2P-VAE to obtain predictions of crush model and Mie-Gr\"uneisen EoS parameters.
\begin{figure}[!htb]
    \centering
    \includegraphics[width=0.4\textwidth]{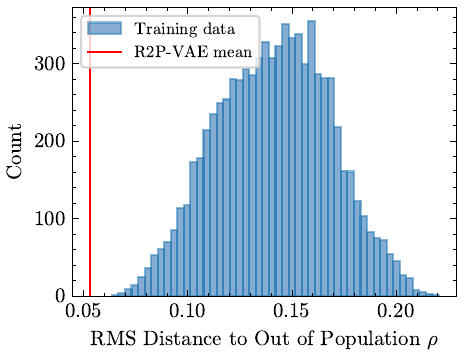}
    \caption{Histogram of the root mean squared (RMS) distance between the density fields for the mismatched EoS test case (visualized in \autoref{fig:sesame_visualization}), and the density fields used to generate the radiographs for training the R2P-VAE. The RMS distance is computed using the density fields for both the high and low impact velocity. The red vertical line indicates the RMS distance between the mean of the R2P-VAE reconstructions and the ground truth. The units of the RMS distance are $\si{g/cm^3}$.}
    \label{fig:sesame_dist_to_train_set}
\end{figure}

\begin{figure}[!htb]
    \centering
    \includegraphics[width=0.9\linewidth]{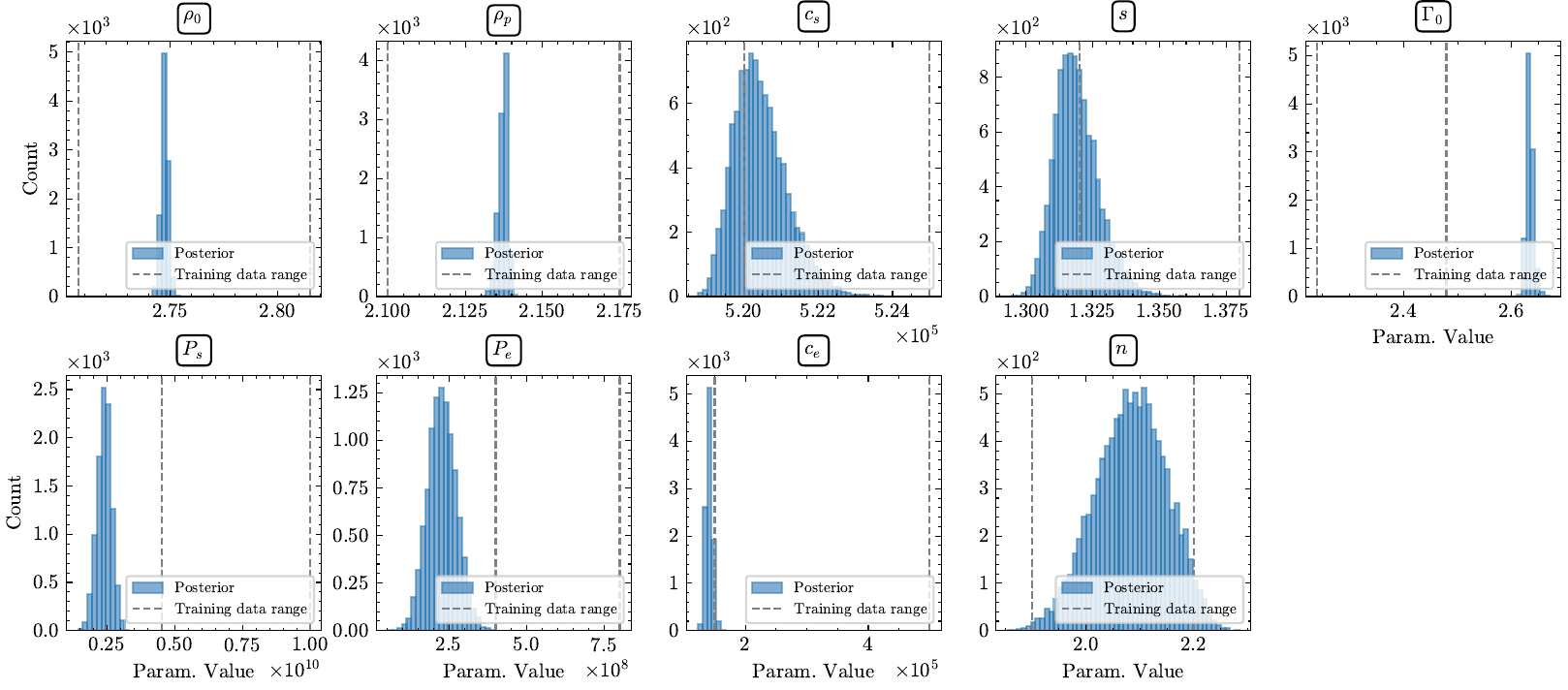}
    \caption{Predicted posterior distribution of parameters for the mismatched EoS test case (reconstructions in \autoref{fig:sesame_visualization}). For each parameter, we show a histogram of 10,000 samples from the posterior. In each subplot, we also indicate the range of the parameter value used for training the R2P-VAE (dashed vertical lines).}
    \label{fig:sesame_posterior}
\end{figure}
In \autoref{fig:sesame_visualization}, we show the density reconstructions obtained using these predicted parameters. Qualitatively, the density reconstructions are reasonable, with the errors most concentrated around the material interfaces and shock features. {Quantitatively, the R2P-VAE reconstructions attain an RMSE of $5.31\cdot10^{-2}~\si{g/cm^3}$ and an MAE of $9.08\cdot10^{-3}~\si{g/cm^3}$ when compared to the true density fields. To better contextualize these results, we computed the root mean squared (RMS) distance between the density fields used to create the training dataset and the density fields produced with the Sesame EoS. In \autoref{fig:sesame_dist_to_train_set}, we plot a histogram of these distances as well as the distance between the mean of the R2P-VAE reconstructions and the ground truth. We find that the R2P-VAE provides a reconstruction that is closer to the out-of-population density fields than any of the 8,000 examples in the training set.}

{We also display the posterior distribution of crush model and Mie-Gr\"uneisen EoS parameters predicted for this test case in \autoref{fig:sesame_posterior}. We show the distribution of predicted parameters, as well as the range of parameter values used for training the network. For many of the parameters, the posterior samples lie partially or entirely outside of the range used for training. This result shows that the R2P-VAE may have some ability to implicitly capture physics in unseen regions of parameter space, since these parameter predictions produce reconstructions that are closer to the true density fields than any of the training examples. We note, however, that the success of this approach also depends on the physical models used for training being expansive enough to closely approximate the unseen physics. Indeed, we remark that the ability of the Mie-Gr\"uneisen EoS to represent the Sesame EoS in this simulation regime is not too surprising, given that the thermodynamic variables are well within the range of applicability of the Mie-Gr\"uneisen EoS, which is typically assumed when the compression of the shocked material is less than about 16\% and there is not a significant release.
% The compression less than ~16 percent and there is not significant release.

% This is somewhat remarkable, since \autoref{fig:sesame_dist_to_train_set} demonstrates that these predictions produce density fields that are more consistent with the ground truth than any training example. This result shows that the R2P-VAE may have some ability to implicitly recognize density fields corresponding to parameter combinations not seen during training. This finding is surprising, but encouraging, and suggests that an additional advantage of our learning based approach may be that it has some ability to generalize to regions of parameter space outside of the training distribution.}

\section{Discussion}
\label{sec:discussion}

In this paper, we introduce a machine learning method for generative prediction of physical parameters from {density fields or} radiographic data using a conditional variational autoencoder. We show that the proposed D2P-VAE can be used to inform the design of flyer plate impact experiments. In particular, we use ML as a tool to demonstrate with high confidence that using only high impact velocity data does not provide adequate information, even with fully resolved density fields or a dynamic sequence of images, to accurately infer EoS and crush model parameters. In contrast, adding an accompanying ``low'' impact velocity experiment that captures a different regime, together with a high impact flyer plate to capture shock propagation allows robust parameter inference of all EoS and crush model parameters. We subsequently apply the R2P-VAE to the synthetic radiographic data, and demonstrate  accurate parameter prediction and reliable uncertainty estimation for many of the equation of state and crush model parameters.   However, we note that the radiographic images result in an information loss relative to the full density fields that degrades parameter estimation. We also demonstrate that combining our parameter estimation framework with traditional hydrodynamic solvers allows for highly accurate density reconstruction, with results that are qualitatively and quantitatively comparable to those obtained with a deep network trained specifically for density reconstruction. Most importantly, the density estimates obtained by propagating the recovered parameters are consistent with the continuous PDEs and underlying conservation laws, unlike traditional reconstruction techniques. {Finally, we demonstrate the effectiveness of the proposed approach under distribution shifts, particularly to data with out-of-distribution radiographic noise and data generated with previously unseen physics. Overall, our results demonstrate that direct parameter estimation from radiographic data could serve as a practical and effective experimental analysis technique.}

{Finally, we remark that the proposed method offers several distinct advantages over existing approaches. In particular, the proposed approach eliminates the need to obtain density reconstructions from experimental radiographs before performing parameter estimation, which is a major obstacle to applying many existing methods. Moreover, the proposed approach enables very fast inference of parameter values, which generally cannot be offered by methods like MCMC or PDE constrained optimization, even if they could be used. We also note that the proposed approach of performing density reconstruction using hydrodynamic solvers after estimating parameters guarantees that the reconstructions will be physically admissible, which is not guaranteed by traditional radiographic inversion methods. Future extensions of the present work could include applying the approach to real experimental data or investigating how we can further increase the approach's robustness to physical and radiographic model mismatches.}

% {The proposed approach eliminates the need to obtain density reconstructions from experimental radiographs before performing parameter estimation, which is a major obstacle to applying many existing methods.} We applied the approach to simulated flyer plate impact experiments and found that the proposed method enables accurate parameter prediction and reliable uncertainty estimation for many of the relevant equation of state and crush model parameters. 

\bibstyle{plain}
\bibliography{ref.bib}

\section*{Acknowledgments}

This work was supported by the Laboratory Directed Research and Development program at Los Alamos National Laboratory. The authors would like to acknowledge Oleg Korobkin for initially developing the script used to compute the Abel transform of the density fields, Soumi De for performing preliminary investigations on parameter estimation using a similar dataset, and Matt Hudspeth for discussions on shocks in porous material.

\section*{Author contributions statement}

E.B., D.A.S., B.S.S., and M.K. conceived the proposed parameter estimation approach. E.B. implemented the machine learning methods. D.A.S., M.K., and T.W. performed the dataset generation and density reconstruction using the hydrocode. D.A.S. prepared Fig. \ref{fig:initial}, while E.B. produced the remaining figures. B.S.S. provided feedback on the results and significantly revised the entire manuscript. M.K. supervised the work, wrote \autoref{sec:intro}, and revised the manuscript. All authors reviewed the final document.

\section*{Competing interests}

The authors declare no competing interests.

\section*{Data availability statement}

The data and code used to generate the results in this study are not currently publicly available, but may be obtained from the authors upon reasonable request.

\appendix

\section{Radiographic system model}
\label{app:radiography}

The flyer plate impact simulations used in our experiments were carried out in two dimensions. These simulations can be used to model real three-dimensional experiments under the assumption that the three-dimensional density field is axially symmetric. We adopt a parallel beam geometry for the X-ray imaging system. Under these assumptions, the \textit{areal density} $\rho_A$ of every material in the experimental apparatus can be obtained by simply applying a forward Abel transform to the density field of each material $\rho_i$, where $i$ indexes all materials. That is, the individual areal densities are $\rho_{A,i} = \mathcal{A}(\rho_i)$, where $\mathcal{A}$ represents the Abel transform.

For an X-ray source with intensity $I_0$, the attenuation of the  radiation is modeled by the Beer-Lambert law as:

\begin{equation*}
    I = I_0 \cdot \exp\left({-\sum_i \xi_i\rho_{A,i}}\right),
\end{equation*}

where $I$ represents the X-ray intensity at the detector and $\xi_i$ is the mass attentuation coefficient of material $i$. This intensity yields the direct radiograph $d$ up to a constant factor $C$, which accounts for factors such as pixel size and exposure time, so that $d = C\cdot I$. We then model the noisy transmission radiograph $T$ as $T = d +n$, where $n$ represents noise from multiple sources.

In particular, we write:

\begin{equation*}
    n = D_{dsb} + D_s + B_s + \eta.
\end{equation*}

$D_{dsb}$ models blur from both the source and detector as $D_{dsb} = \phi_{db} * G_{blur}(\sigma_{blur}) * d$, where $\phi_{db}$ is a hand-designed detector blur kernel and $G_{blur}(\sigma_{blur})$ is a 2D Gaussian kernel with standard deviation $\sigma_{blur}$. $D_s$ represents correlated scatter, calculated as $D_s = \kappa \cdot G_{scatter}(\sigma_{scatter}) * d$, where $\kappa$ is a constant and $G_{scatter}(\sigma_{scatter})$ is another 2D Gaussian kernel. $B_s$ is an uncorrelated tilted background scatter field, which is given by $B_s = ax + by$, where $a$ and $b$ are constants and $x$ and $y$ are coordinates in the detector plane. Finally, we add correlated gamma and photon noise $\eta$, which is modeled as $\eta = \phi_g * \text{Pois}(\gamma_g) + \phi_p * \text{Pois}(\gamma_p)$, where $\gamma_g$ and $\gamma_p$ are the signal-dependent rates, and $\phi_g$ and $\phi_p$ are the gamma and photon kernels.

\section{VAE architecture and implementation}
\label{sec:AppR2PVAE}

\begin{figure}[!htb]
    \centering
    \includegraphics[]{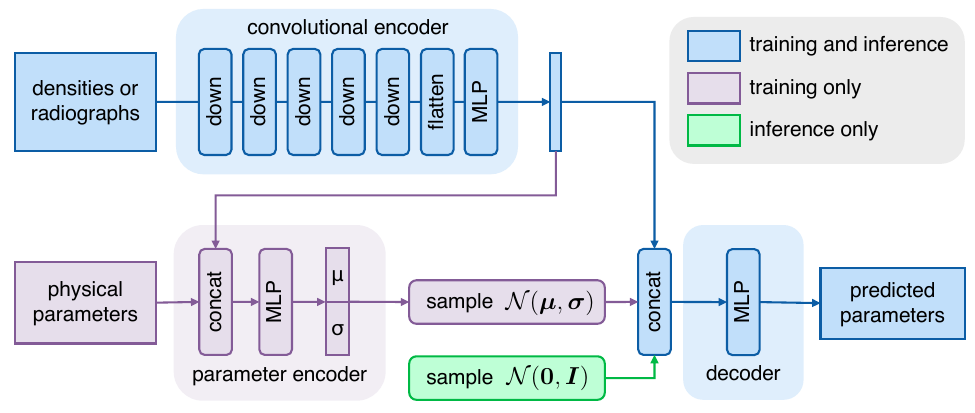}
    \caption{Architecture of the proposed D2P/R2P-VAE. Inputs and components used during both training and inference are colored in blue. Inputs and components used during training only (such as the true parameters and the parameter encoder) are colored purple. Components used during inference only are marked in green. The internal structure of the convolutional ``down'' blocks is shown in \autoref{fig:up_and_down_blocks}.}
    \label{fig:r2pvae_flowchart}
\end{figure}

\begin{figure}[!htb]
    \centering
    \includegraphics[]{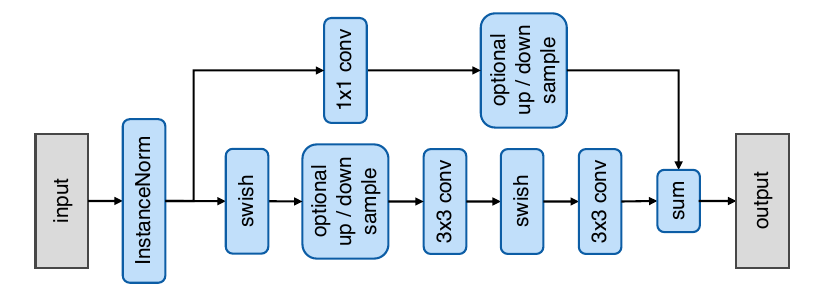}
    \caption{Internal structure of the convolutional blocks used in the D2P/R2P-VAE and R2D-Net.}
    \label{fig:up_and_down_blocks}
\end{figure}

The D2P/R2P-VAE is a variational autoencoder that is conditioned directly on images, which are either density fields or radiographs. The images are first passed through a convolutional encoder to produce a compressed representation. At training time, this representation is used by both the parameter encoder and decoder. During training, the physical parameters and compressed images are passed through a multi-layer perceptron (MLP) parameter encoder that outputs the mean $\bmu$ and standard variation $\bsigma$ of a multi-variate normal distribution. A sample from $\mathcal{N}(\bmu, \bsigma)$ is then concatenated with the compressed image representation and passed through an MLP decoder to reconstruct the parameters. A schematic of the proposed network architecture is given in \autoref{fig:r2pvae_flowchart}.

The loss function used for training VAEs is a combination of two terms: a reconstruction error term, which ensures that that the VAE approximately acts as an autoencoder, and a Kullback-Leibler (KL) divergence term that ensures the learned latent space is close to a standard normal distribution, so that passing samples of the $\mathcal{N}(\bzero, \bI)$ prior through the decoder approximates the posterior \cite{VAE}. Since accurate reconstruction requires informative latent representations, these two objectives are at odds with each other. Ensuring the correct balance between these terms is important for producing high-quality samples and avoiding the common ``posterior collapse'' phenomenon\cite{posteriorcollapse}, where the VAE's predictions collapse to a point estimate of the posterior mean.

The most common method of balancing these objectives is the $\beta$-VAE \cite{betaVAE}. Explicitly, for parameters $\p$ and reconstructed parameters $\hat{\p}$, the loss is given by:

\begin{equation}
    \label{eq:beta_VAE_loss}
    \mathcal{L}_{\beta\text{-VAE}} = \frac{D}{2} \text{MSE}(\p, \hat{\p}) + \beta \cdot \mathcal{D}_{\text{KL}} \infdivx{\mathcal{N}(\bmu, \bsigma)}{\mathcal{N}(\bzero, \bI)},
\end{equation}

where $D$ is the dimension of $\p$, $\text{MSE}$ denotes the mean squared error and $\mathcal{D}_{\text{KL}}$ is the KL divergence. While often effective, the $\beta$-VAE suffers from two primary drawbacks: (1) finding a good value of $\beta$ requires manual tuning, which may be expensive, and (2) the value of the $\beta$-VAE objective loses a rigorous statistical interpretation as the evidence lower bound (ELBO), which is the basis of the original VAE.

To address both of these issues, we adopt the recently proposed $\sigma$-VAE method \cite{sigmavae}. The $\sigma$-VAE re-examines the implicit assumption of the $\beta$-VAE loss function, which is that the decoder output $\hat{\p}$ is the mean of a distribution of admissible reconstructions $\mathcal{N}(\hat{\p}, \bI)$. Minimizing the negative log likelihood of the real data then leads to the MSE loss in \autoref{eq:beta_VAE_loss}. In its simplest form, the $\sigma$-VAE instead uses the calibrated distribution $\mathcal{N}(\hat{\p}, \sigma^2 \bI)$. Summing the negative log likelihood and KL divergence terms then yields the $\sigma$-VAE loss:

\begin{equation}
    \label{eq:sigma_VAE_loss}
    \mathcal{L}_{\sigma\text{-VAE}} = D \log \sigma + \frac{D}{2\sigma^2} \text{MSE}(\p, \hat{\p}) + \mathcal{D}_{\text{KL}} \infdivx{\mathcal{N}(\bmu, \bsigma)}{\mathcal{N}(\bzero, \bI)}.
\end{equation}

Optimizing the $\sigma$-VAE objective is equivalent to optimizing the $\beta$-VAE objective, if one fixes $\sigma^2 = \beta$. However, instead of manually choosing the value of $\sigma$, the $\sigma$-VAE estimates the optimal value of $\sigma$ for every batch. This estimate is simply given by $(\sigma^*)^2 = \mathbb{E}_\p [\mathbb{E}_{\hat{\p}} [\text{MSE}(\p, \hat{\p})) ]]$, where the inner expectation over $\hat{\p}$ implicitly captures the entire encoding and decoding process, which is also conditioned on the noisy radiographs. The optimality of this choice follows from the fact that the maximum likelihood estimate of the variance is the average squared distance to the mean, i.e. the expected value of $\text{MSE}(\p, \hat{\p})$. In practice, we approximate the inner expectation with a single sample of $\hat{\p}$ per data point and the outer expectation with a batch of data points. This is the approach suggested in\cite{sigmavae}, where it was shown that this sampling technique introduces an inconsequential error in estimating $\sigma^*$ while avoiding additional computational costs. Our complete training objective is then given by \autoref{eq:sigma_VAE_loss}, with $\sigma$ replaced by the $\sigma^*$ estimated from each batch. In practice, we found that the $\sigma$-VAE produced good results with little manual hyperparameter tuning.

\subsection{Training and implementation details}

The convolutional encoder of the D2P/R2P-VAE consists of 5 downsampling blocks. We show the complete internal structure of each of these blocks in \autoref{fig:up_and_down_blocks}. The normalization layer used in these blocks is InstanceNorm\cite{instancenorm}, and the activation function is the sigmoid linear unit, also referred to as SiLU or swish\cite{swish}. The downsampling operation is $2\times2$ average pooling. The input to the convolutional encoder is the relevant image (or images, concatenated along the channel dimension). The first downsampling block increases the channel dimension to eight, and each subsequent block adds another eight channels. The output of the last downsampling block is then flattened into a vector, and passed through a feedforward network which also uses a SiLU activation. The input and subsequent activations of this feedforward network have dimension $(520 \cdot \text{number of input channels} \to 256 \to 256)$.

The encoder of the VAE takes the true parameters and the compressed representation of the radiograph concatenated together as input. The encoder is a feedforward network with SiLU activation. The input and subsequent activations have dimensions $(265 \to 256 \to 128)$. This $128$ dimensional vector is then split into 64 dimensional vectors $\bmu$ and $\bsigma$. The compressed radiograph is concatenated with a sample from $\mathcal{N}(\bmu, \bsigma)$ and passed through the decoder. The decoder is another SiLU feedforward network, with input and activation dimensions of $(320 \to 256 \to 9)$.

During training, the network weights were updated using the AdamW optimizer\cite{adamw} with a learning rate of $1\cdot10^{-4}$. To improve numerical stability, we follow a suggestion of the original $\sigma$-VAE implementation and \textit{softclip} the entires of $\bsigma$ and the value of $\sigma^*$ to a minimum of $e^{-4}$, where the softclip function is defined by $\text{softclip}_m(x) := m + \log( 1 + e^{x-m})$, where $m$ is the minimum allowed value. All models were implemented with PyTorch\cite{paszke2019pytorch} and trained using PyTorch Lightning\cite{pytorch_lightning}. We split the data randomly into 8,000 cases for training, 1,000 cases for validation, and 1,000 cases for testing. We trained the D2P/R2P-VAE on a computer with eight NVIDIA GeForce RTX 2080 Ti GPUs. The models typically achieved their minimum validation losses after between six and twelve hours of training.

Because the values of the physical parameters we are estimating have extremely different absolute scales, it is beneficial to normalize the values of each parameter during training. We apply $z$-score standardization to each parameter, where the mean and standard deviation of the parameter are computed online during training using Welford's algorithm\cite{welfordsalgorithm}. The mean and standard deviation of the parameter values used for normalization are then fixed during validation and testing.

\section{Parameter estimates from radiographs}
\label{app:CsAndS}

\begin{figure}[!htb]
    \centering
    \includegraphics[width=0.95\linewidth]{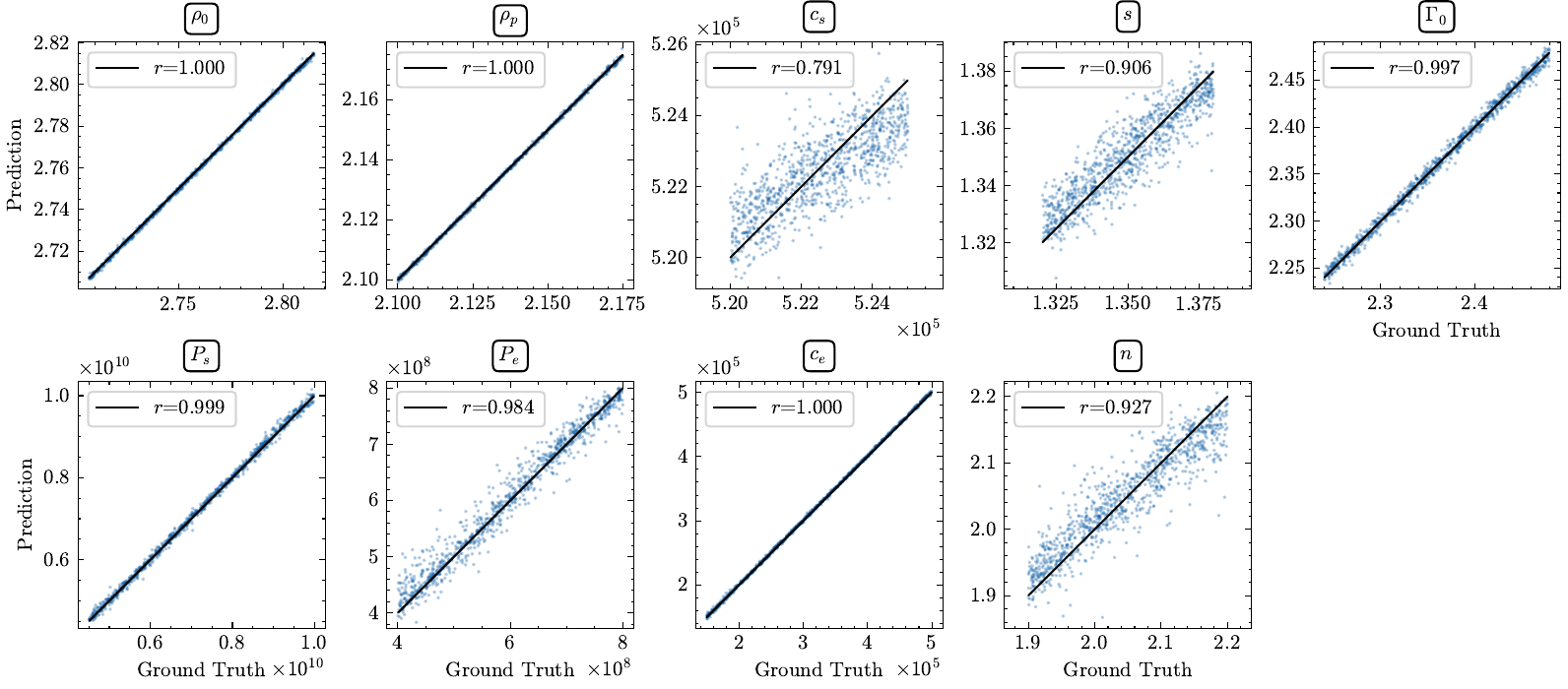}
    \caption{Parameter estimates from the trained R2P-VAE on the testing set with \textit{clean} radiographs used as input. For each test case, we use the mean of 1,000 posterior samples as a point estimate of the parameter values. We also report the Pearson correlation coefficient $r$ in each subplot. Compare with estimates from noisy radiographs in \autoref{fig:MMSE_param_estimates_new}.}
    \label{fig:clean_rads_correlations}
\end{figure}

\begin{figure}[!htb]
    \centering
    \includegraphics[width=\linewidth]{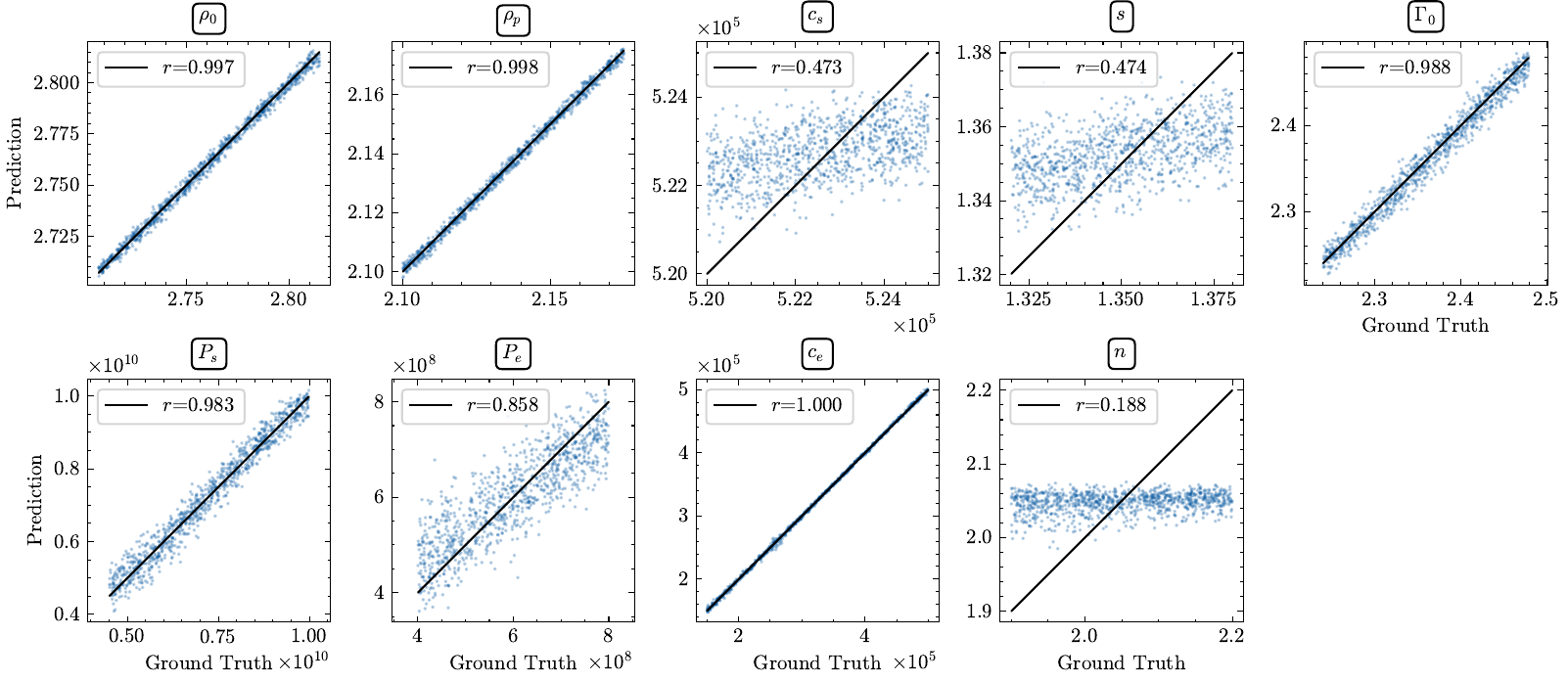}
    \caption{Parameter estimates from the trained R2P-VAE on the testing set with \emph{noisy} radiographs used as input. For each test case, we use the mean of 1,000 posterior samples as a point estimate of the parameter values. We also report the Pearson correlation coefficient $r$ in each subplot. Compare with estimates from clean radiographs in \autoref{fig:clean_rads_correlations}.}
    \label{fig:MMSE_param_estimates_new}
\end{figure}

\section{R2D-Net}
\label{app:R2D-Net}

The radiographs-to-density network (R2D-Net) is convolutional neural network based on the U-Net \cite{unet} architecture. For a fair and complete comparison with the R2P-VAE, the R2D-Net accepts a pair of noisy radiographs as input, and is trained to output both corresponding density fields.

\begin{figure}[!htb]
    \centering
    \includegraphics[]{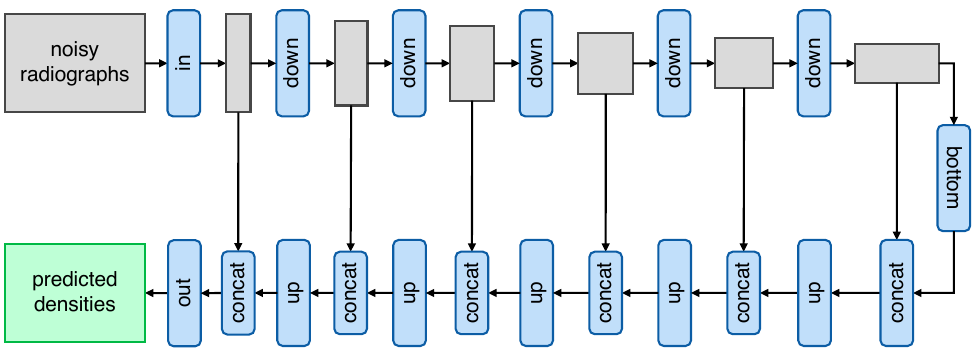}
    \caption{Architecture of the R2D-Net. The internal structure of the convolutional blocks is shown in \autoref{fig:up_and_down_blocks}. The blocks labeled ``in,'' ``out,'' and ``bottom'' have no upsampling or pooling.}
    \label{fig:r2dnet_arch}
\end{figure}

The complete architecture of the R2D-Net is shown in \autoref{fig:r2dnet_arch}. The architecture has five downsampling blocks and five upsampling blocks, which are very similar to the residual blocks found in BigGAN \cite{biggan}, and identical to those found in \cite{deblurring_via_sr}. The complete internal structure of these blocks is displayed in \autoref{fig:up_and_down_blocks}. The ``in'' block increases the number of image channels from two (one for each experiment) to 64. In each subsequent downsampling block, the number of channels increases by 64, while the spatial dimensions are downsampled by a factor of two. The upsampling path reverses this process, so that the number of channels decreases by 64 after each upsampling block and the spatial dimensions are upsampled by a factor of two. The downsampling operation used is $2\times2$ average pooling, and the upsampling operation is $2\times2$ nearest neighbor upsampling. In the up and down blocks, the first of the $3\times3$ convolutions increases/decreases the number of channels. As is characteristic of U-Net architectures, we use skip connections to combine features from the downsampling path with features in the upsampling path. Feature maps of the same spatial dimensions are concatenated together along the channel dimension before they are passed through each upsampling block.

The R2D-Net was trained to minimize the mean squared error of the reconstructed density fields. During training, the model weights were updated using the AdamW optimizer with a learning rate of $1\cdot10^{-4}$ and a batch size of 32. The training and testing split that was used for the R2P-VAE was also used for training and evaluating the R2D-Net. We trained the R2D-Net on a computer with eight NVIDIA GeForce RTX 2080 Ti GPUs, and the model achieved its minimum validation loss after approximately 11 hours of training.

\subsection{Density reconstruction results}

{Representative density reconstructions from the trained R2D-Net are shown in \autoref{fig:median_r2d_test_recon}. Unlike the R2P-VAE, the R2D-Net only provides a single estimate of the density fields for every test case. We find that, like the R2P-VAE, the R2D-Net provides highly accurate density reconstructions that are very close to the ground truth.}

\begin{figure}[!htb]
    \centering
    \includegraphics[width=0.8\textwidth]{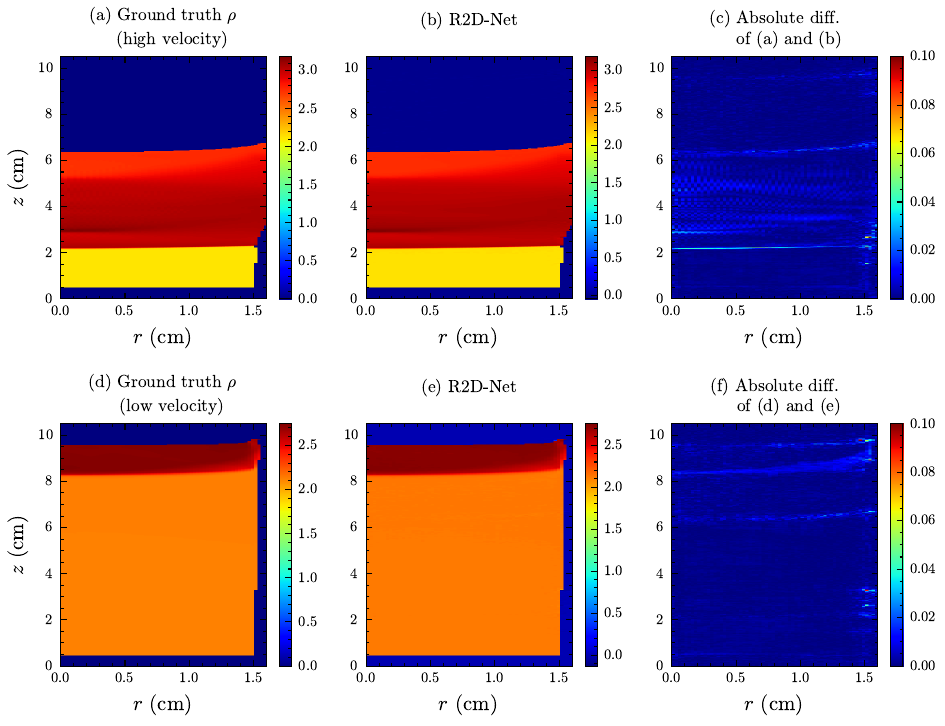}
    
    \caption{Density reconstructions directly from noisy radiographs using the trained R2D-Net for a representative test case. The top row of figures corresponds to the high impact velocity, and the bottom row corresponds to the low impact velocity. Subfigures: (a) ground truth density field, (b) predicted density from noisy radiographs, (c) absolute difference of (a) and (b); subfigures (d-f) display the corresponding reconstructions at the lower impact velocity. Note the scale of the colorbars in (c) and (f), which have been truncated at 0.1 g/cm$^3$ for visibility.}
    \label{fig:median_r2d_test_recon}
\end{figure}

{We quantify the accuracy of the reconstructions obtained with the R2D-Net and compare them to the reconstructions obtained with the R2P-VAE in \autoref{tab:error_table_both}. We computed the RMSE and MAE for both methods across the testing set of 1,000 cases. We also computed the metrics over the regions of interest shown in \autoref{fig:roi}. We find that the R2D-Net outperforms the R2P-VAE in terms of RMSE of the entire recovered density fields in both the in-distribution and OOD noise settings. This finding is expected: the R2D-Net is optimized to minimize the MSE of the reconstructed density fields, so it should perform the best in terms of this metric. On the other hand, we find that the R2P-VAE outperforms the R2D-Net in terms of MAE across both experimental settings. We believe that this is because the R2P-VAE approach has a tendency to produce large errors in a very small number of pixels, especially along the material interfaces (see \autoref{fig:median_r2p_test_recon}). These types of errors are much more heavily penalized by MSE than MAE.}

\begin{table}[!htb]
    \centering
    \begin{tabular}{ccccc}
    \toprule
     &  & R2P-VAE mean & R2P-VAE samples & R2D-Net \\ \midrule
    \multirow{2}{*}{RMSE} & In-Population & $1.98\cdot 10^{-2}$ & $2.71\cdot 10^{-2}$ & $\mathbf{6.96\cdot 10^{-3}}$   \\
     & OOD Noise & $3.37\cdot 10^{-2}$ & $3.93\cdot 10^{-2}$ & $\mathbf{3.16\cdot 10^{-2}}$   \\ \midrule
     
     \multirow{2}{*}{MAE} & In-Population & $\mathbf{1.84 \cdot 10^{-3}}$ & $2.20\cdot 10^{-3}$ & $2.48\cdot 10^{-3}$   \\
     & OOD Noise & $\mathbf{4.42\cdot 10^{-3}}$ & $4.66\cdot 10^{-3}$ & $8.15\cdot 10^{-3}$  \\ \midrule

    \multirow{2}{*}{Mean $L^\infty$ Error} & In-Population & $2.203$ & $2.818$ & $\mathbf{0.851}$ \\
     & OOD Noise & $\mathbf{2.743}$ & $2.944$ & $2.814$ \\ \midrule
     
     \multirow{2}{*}{RMSE over ROI} & In-Population & $5.96 \cdot 10^{-3}$ & $7.91\cdot 10^{-3}$ & $\mathbf{5.22\cdot 10^{-3}}$   \\
     & OOD Noise & $\mathbf{1.36\cdot 10^{-2}}$ & $1.46\cdot 10^{-2}$ & $1.48\cdot 10^{-2}$  \\ \midrule
     \multirow{2}{*}{MAE over ROI} & In-Population & $\mathbf{3.64 \cdot 10^{-3}}$ & $4.67\cdot 10^{-3}$ & $3.69\cdot 10^{-3}$   \\
     & OOD Noise & $\mathbf{7.61\cdot 10^{-3}}$ & $8.33\cdot 10^{-3}$ & $1.02\cdot 10^{-2}$ \\ \midrule
    \multirow{2}{*}{Mean $L^\infty$ Error over ROI} & In-Population & $4.64\cdot10^{-2}$ & $6.34\cdot 10^{-2}$ & $\mathbf{3.84\cdot10^{-2}}$\\
     & OOD Noise & $1.30\cdot10^{-1}$ & $1.38\cdot10^{-1}$ & $\mathbf{1.21\cdot10^{-1}}$ \\
     \bottomrule
    \end{tabular}
    \caption{Table of root mean squared error (RMSE) and mean absolute error (MAE) of density reconstruction methods for both in-population radiographic data and radiographs corrupted with out-of-distribution (OOD) noise. We also report the metrics computed over the regions of interest displayed in \autoref{fig:roi}.}
    \label{tab:error_table_both}
\end{table}

{Finally, we also compare the R2D-Net and R2P-VAE reconstructions in the presence of mismatched physics. We passed the radiographs for the test case generated with a Sesame EoS (shown in \autoref{fig:sesame_visualization}) through the R2D-Net to obtain density reconstructions. In \autoref{tab:errors_sesame}, we compare these reconstructions quantitatively with the R2P-VAE reconstructions. We find that R2D-Net achieves a lower RMSE and MAE than the R2P-VAE for this test case, although the gap is not very large. This result is also somewhat unsurprising. We speculate that since the R2D-Net produces density fields directly, it may be less inhibited by the EoS mismatch than the R2P-VAE approach, which implicitly requires that the EoS used for training can accurately approximate the mismatched EoS for the density reconstructions to be accurate.}

\begin{table}[!htb]
    \centering
    \begin{tabular}{cccc}
    \toprule
     & R2P-VAE mean & R2P-VAE samples & R2D-Net \\ \midrule
     RMSE & $5.31\cdot 10^{-2}$ & $6.04\cdot 10^{-2}$ & $\mathbf{2.15\cdot 10^{-2}}$   \\ \midrule
     
     MAE & $9.08 \cdot 10^{-3}$ & $9.22\cdot 10^{-3}$ & $\mathbf{7.73\cdot 10^{-3}}$   \\
     \bottomrule
    \end{tabular}
    \caption{Table of root mean squared error (RMSE) and mean absolute error (MAE) of density reconstruction methods for a test case generated with mismatched physics (a Sesame EoS instead of the Mie-Gr\"uneisen EoS).}
    \label{tab:errors_sesame}
\end{table}

\section{Additional figures and tables}
\label{app:additional_figs}

\begin{figure}[!htb]
    \centering
    \includegraphics[width=0.6\textwidth]{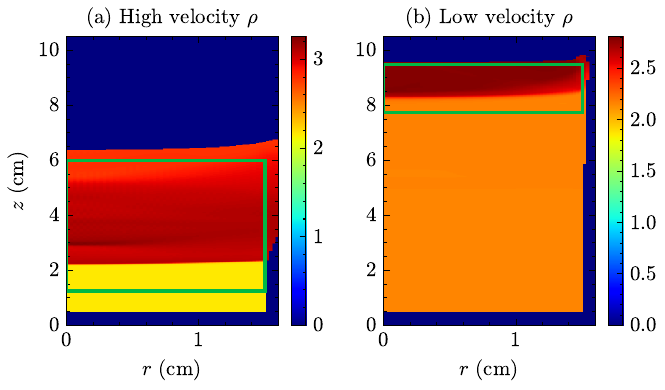}
    \caption{Green boxes bound the regions of interest used to compute the ROI metrics in \autoref{tab:error_table_r2p} and \autoref{tab:error_table_both}. Subfigure (a) shows the ROI for an example density field from the high impact velocity experiment, whereas subfigure (b) depicts the low impact velocity experiment.}
    \label{fig:roi}
\end{figure}

\begin{table}[!htb]
\centering
\begin{tabular}{cccccccccc}
\toprule
& $\rho_0$ & $\rho_p$  & $c_s$ & $s$ & $\Gamma_0$ & $P_s$ & $P_e$  & $c_e$  & $n$ \\
\midrule
\autoref{fig:initial} & 2.713 & 2.133 & $5.223\cdot10^5$  & 1.368  & 2.368 & $6.514\cdot10^9$ & $6.085\cdot10^8$ & $2.721\cdot10^5$ & 1.995 \\
\autoref{fig:density_rad_feat_vis} & 2.771 & 2.127 & $5.206\cdot10^5$  & 1.352  & 2.274 & $8.765\cdot10^9$ & $5.421\cdot10^8$ & $2.137\cdot10^5$ & 1.950 \\
\autoref{fig:line_outs} & 2.761 & 2.138 & $5.500\cdot10^5$  & 1.500  & 2.360 & $7.250\cdot10^9$ & $6.000\cdot10^8$ & $3.250\cdot10^5$ & 2.05 \\
\autoref{fig:line_outs_varied_cs_s} & 2.761 & 2.138 & various & various  & 2.360 & $7.250\cdot10^9$ & $6.000\cdot10^8$ & $3.250\cdot10^5$ & 2.05 \\
\autoref{fig:posteriors_new} & 2.771 & 2.127 & $5.206\cdot10^5$  & 1.352  & 2.274 & $8.765\cdot10^9$ & $5.421\cdot10^8$ & $2.137\cdot10^5$ & 1.950 \\
\autoref{fig:median_r2p_test_recon} & 2.771 & 2.127 & $5.206\cdot10^5$  & 1.352  & 2.274 & $8.765\cdot10^9$ & $5.421\cdot10^8$ & $2.137\cdot10^5$ & 1.950 \\
\autoref{fig:ood_noise} & 2.719 & 2.143 & $5.233\cdot10^5$  & 1.345  & 2.306 & $9.490\cdot10^9$ & $6.713\cdot10^8$ & $3.810\cdot10^5$ & 2.124  \\
\autoref{fig:median_r2d_test_recon} & 2.768 & 2.164 & $5.209\cdot10^5$  & 1.337  & 2.401 & $6.171\cdot10^9$ & $7.715\cdot10^8$ & $3.993\cdot10^5$ & 1.919 \\
\midrule
units & [$g/cm^3$] & [$g/cm^3$] & [$cm/s$] & -- & -- & [$dyne/cm^2$] & [$dyne/cm^2$] & [$cm/s$] & -- \\
\bottomrule
\end{tabular}
\caption{Parameter values used in figures throughout the manuscript.}
\label{tab:figure_params}
\end{table}

\end{document}